\documentclass[preprint,showpacs,preprintnumbers,amsmath,amssymb]{revtex4}

\usepackage{graphicx}
\usepackage{dcolumn}
\usepackage{bm}

\def\br{\left(\begin{array}{c}}
\def\er{\end{array}\right)}

\def\sinW2{\sin^2\theta_W}

\def\mz2{M_{z}^2}
\def\c2b{\cos 2\beta}

\def\c2{{\tilde\chi^{+}_2}}

\def\mz{M_z}

\def\Fq2{F_{2}(q^2)}
\def\f{\({\cal F}\)}
\def\d1{{\f(\tilde c;\tilde s;\tilde W)+ \f(\tilde c;\tilde \mu;\tilde W)}}

\def\sec2w{sec^2\theta_W}

\def\r2{\sqrt 2}
\def\beq{\begin{equation}}
\def\eeq{\end{equation}}
\def\beqn{\begin{eqnarray}}
\def\eeqn{\end{eqnarray}}

\def\sinW2{\sin^2\theta_W}

\def\mz2{M_{z}^2}
\def\c2b{\cos 2\beta}

\def\mz{M_z}

\def\Fq2{F_{2}(q^2)}
\def\f{\({\cal F}\)}
\def\d1{{\f(\tilde c;\tilde s;\tilde W)+ \f(\tilde c;\tilde \mu;\tilde W)}}

\def\sec2w{sec^2\theta_W}

\def\c2{{\tilde\chi^{+}_2}}

\def\mz{M_z}

\def\Fq2{F_{2}(q^2)}
\def\f{\({\cal F}\)}
\def\d1{{\f(\tilde c;\tilde s;\tilde W)+ \f(\tilde c;\tilde \mu;\tilde W)}}

\def\sec2w{sec^2\theta_W}

\begin{document}

\preprint{CERN-PH-TH/2008-139}

\title{An MSSM Extension with a  Mirror Fourth Generation,  Neutrino Magnetic  Moments and LHC Signatures}

\author{Tarek Ibrahim$^{a,b}$}
 \email{tarek@lepton.neu.edu}
\author{Pran Nath$^{b,c}$}%
 \email{nath@lepton.neu.edu}
\affiliation{%
a. Department of  Physics, Faculty of Science, University of Alexandria, Alexandria, Egypt\\ 
b. Department of Physics, Northeastern University,
Boston, MA 02115-5000, USA\\
c. TH Division, PH Department, CERN, CH-1211 Geneva 23, Switzerland} 


\date{\today}

\begin{abstract}
Recent analyses have shown that  a sequential fourth generation can be consistent with
precision electroweak data. We consider the possibility that the new generation could be
a mirror generation with $V+A$ rather than $V-A$ interactions. Specifically we consider
an extension of the minimal supersymmetric standard model with a light mirror generation 
(mirMSSM) . Implications of this extension are 
explored. One consequence is an enhancement of the tau neutrino magnetic moment by
 several orders of magnitude consistent with the current limits on the magnetic moment of the tau.
The masses of the mirror generation arise due to electroweak symmetry breaking, and 
if  a mirror generation exists its mass spectrum must lye below a TeV, and thus should be
discovered at the LHC. Mirror particles  and  mirror sparticles produce many characteristic signatures which should be detectable at the LHC.  Heavy higgs boson decays into mirror particles and 
an analysis of the forward-backward asymmetries can  distinguish a mirror generation from a 
 sequential fourth generation. The validity of the model  can  thus be tested at the LHC.  A  model of the type discussed here could arise from a more unified structure such as grand unification or strings where a mirror generation escapes the survival hypothesis, i.e., a generation and a mirror generation
do not tie up to acquire a mass of size $M_{GUT}$ or $M_{string}$ due to a symmetry,
and thus remain massless down to the electroweak scale.
\end{abstract}

\maketitle

\section{Introduction}
Recent investigations have shown  that a fourth generation is not ruled out by the 
precision electroweak  data  if it is heavy with
masses in the few hundred GeV range
(For recent works see \cite{Kribs:2007nz,Hung:2007ak,Holdom,Novikov:2002tk,Dubicki:2003am,Murdock:2008rx,Cakir:2008su}
and for early works see \cite{Hewett:1986uu,Barger:1989dk,Frampton:1999xi}).
These investigations have typically assumed
 that the fourth generation is a sequential generation
with $V-A$ type interactions.  However, 
an intriguing possibility exists that the new generation 
could be a mirror generation with $V+A$ interactions. 
Mirror generations do arise in unified models of fundamental 
interactions\cite{grs,Georgi:1979md,Wilczek:1981iz,Babu:2002ti},  
and thus it is natural that one consider the
existence of a mirror generation.
Normally one assumes the so called survival hypothesis\cite{Georgi:1979md}
 where with
 $n_f$ number of ordinary families and $n_{mf}$ number of mirror families, 
only $n_f-n_{mf}$ (for $n_f>n_{mf}$) remain light, 
 and the remainder acquire GUT or string scale size masses.
 However, this need not always be the case.
Indeed there are  many escape mechanisms where  
residual symmetries in breaking at the string scale or GUT scale will keep some
mirror families light while others 
 become superheavy \cite{Bagger:1984rk,Senjanovic:1984rw}.
 Mixings between ordinary families and mirrors can arise from non-rernormalizable interactions
 after  spontaneous breaking (see, e.g., \cite{Senjanovic:1984rw,Chang:1985jd}).
Additional work on model building using mirrors  can be found in
 \cite{Bars:1980mb,delAguila:1984qs,Maalampi:1988va,mirrors,Adler:2002yg,Chavez:2006he}
and further implications of mirrors are explored in 
\cite{Nandi:1981qr,Langacker:1988ur,Choudhury:2001hs,Csikor:1994jg,Montvay:1997zq,Triantaphyllou:1999uh}.\\

In this work we make the 
specific assumption that  there is indeed a light mirror generation
with masses below the TeV scale which would be accessible
at the LHC. The  assumption of a full mirror generation leaves the theory 
anomaly free. Essentially all of the analyses valid for a sequential fourth generation
regarding consistency with the precision electroweak data and other constraints 
should be valid for a mirror generation and we assume this to be the case. 
The analysis we present here differs from previous works in many respects. 
First we propose an extension of the minimal supersymmetric standard model
 with a  full mirror generation which is light 
(mirMSSM),
 i.e., with masses below
the TeV scale which will be accessible at the LHC.
Such an extension is not considered in any
of the previous works. Indeed most of the previous   analyses are not in supersymmetric
frameworks. Second we assume that the mixings of the mirror generation occur mostly
with the third generation, and are negligible with the first two generations if they 
occur at all. With this assumption, the $V-A$ structure of the weak interactions for the
first two generations remains intact, while the third generation can develop a small
$V+A$ component. Current data on the third generation  do not necessarily rule
out this possibility. \\

If a mirror generation exists, it would be discovered at the LHC 
with the same amount of  luminosity as  for the a sequential fourth generation which
is estimated to be  $50 fb^{-1}$. 
A mirror generation will lead to interesting and even dramatic
multilepton and  jets signatures which can discriminate between 
a mirror generation and a sequential fourth generation. 
Further, tests of the mirror generation can come from the decay of the heavy Higgs and
 via measurements 
of the forward -backward asymmetry.  Another effect of  the mixings of the
mirror generation with the third generation is on  magnetic moments. 
We analyze these in the  leptonic sector in detail and show that the 
tau neutrino magnetic moment is enhanced by several orders of magnitude
beyond what one has in the  Standard Model. 
 We note in passing
that the term mirror  has also been used in an entirely
different context of mirror worlds\cite{okun,Mohapatra:2005ng} 
where one has  mirror matter with their own mirror  gauge group. The analysis here has  no
relationship with those theories. \\

The outline of the rest of the paper is as follows. In Sec.(2)
we present an extension of the minimal supersymmetric standard model (MSSM)
to include a fourth generation which we assume is a mirror generation 
and allow for a   mixing of this generation with
the 3rd generation.  
Here the interactions in the 
charged and neutral current sectors are worked out including the 
supersymmetric interactions involving the mirrors, the chargions
and the neutralinos. 
Further details of mixing and interactions are given in Appendix A.
An analysis of the $\tau$ neutrino magnetic moment is given in Sec.(3).
Here contributions arise from exchanges of the leptons from the
third generation and from the mirror  generation, and also from the
exchanges of the sleptons and mirror sleptons. 
An analysis of the $\tau$-lepton  anomalous magnetic moment
when mixings with the mirror  family are allowed is given 
in Sec.(4) again including exchanges from the 3rd generation leptons and
sleptons  and from the mirror  
leptons and mirror sleptons. A discussion of the constraints on a mirror generation and a 
quantitative 
analysis of the sizes is given in Sec.(5) in the framework of an extended supergravity
unified model\cite{msugra} which includes the mirror sector. 
When compared with the magnetic moment
analyses in MSSM with or without CP violation\cite{susyg2,cpg2,Ibrahim:2007fb} 
one finds that the 
tau neutrino magnetic moment can be orders of magnitude
larger than in the Standard Model while the magnetic moment of the
tau lies within experimental bounds. 
A qualitative
analysis of the signatures of the mirror  generation at the LHC is given in
Sec.(6). Here  it is shown that some characteristic signatures arise,
such as dominance of $\tau$s in the decay patterns of the 
mirror leptons which should allow one  to discriminate this model
from  other  supersymmetric models. Further, we discuss how one may 
distinguish a mirror  generation from a sequential fourth generation. 
Here aside from the leptonic signatures, the decay of the heavy Higgs bosons, and  the
analysis of the forward-backward asymmetry would allow one to discriminate a 
mirror generation from a sequential fourth generation.  Further details of the
decay of heavy Higgs to mirror fermions are given in Appendix B.
 Conclusions are given in Sec.(7).

\section{Extension of MSSM with a Mirror Generation}
The fourth generation which we assume to be mirror will in general mix with 
the other three generations. However, as is the case for the first three generations
the mixings between the generations get smaller as the ratio of the masses get
further apart. Thus, for example, $V_{ub}<< V_{us}$, and we expect a similar 
phenomenon for mixings involving the fourth (mirror) generation, i.e., we expect $V_{uB} <<V_{ub}$
where $B$ is the 4th (mirror) generation bottom quark. 
As an example, the mixing between the first and the second can be 
estimated by the Gatto-Sartori-Tonin-Oakes relation\cite{gsto} $V_{us}= \sqrt{m_d/m_s}$ which gives 
 $V_{us}$ to be about $0.2$.
The mixing of the first with the third can be very roughly estimated so that $V_{ub}=\sqrt{m_d/m_b}$ which gives about 
$.03$, i.e., a factor about 10 smaller than $V_{us}$\footnote{This is actually a significant over estimate since the
most recent CKM fits give a value which is even smaller, i.e., $V_{ub}=(3.93\pm .36)\times 10^{-3}$\cite{pdg}.}.
If we extend this  rough estimate to the fourth generation one will have mixing between the first and the fourth as $V_{uB}= \sqrt{m_d/m_B} = .005 $(for $m_B=200$ GeV). Assuming similar mixings will hold in the leptonic sector one will have mixings between the first and the fourth as $ \sqrt{m_e/m_E} = .0016$ (for $M_E$=200 GeV)
where $E$ is the 4th (mirror) generation lepton.
More detailed analyses using error bars on electroweak data
show that the constraints on the enlarged CKM matrix are more relaxed\cite{Kribs:2007nz} 
(see also Sec.V).  Conversely it means that with the current limits on the mixing angles 
  the effects of the 4th generation on the analysis of the electroweak data
lie well within the error bars. Here the electroweak parameters which require special attention are
the S, T, U variables where larger contributions from the 4th generation are possible, but still the 
data can  be made compatible with a  4th generation. Returning to the mixing of the 4th generation with 
the first two one can easily check that small mixings of the type discussed above 
  lead to negligible effect of the 4th generation on the
phenomenology of the first two generations. For this reason we will make a simplifying assumption
of neglecting the mixing effects of the fourth with the first two generations
and  consider below the mixing of just the third and the fourth. 
However,  the following  analysis can be straightforwardly extended to the full four generations  by
letting the  generation index  run from 1-4 keeping in mind that the 4th  generation is a mirror generation.
Thus under $SU(3)_C\times SU(2)_L \times U(1)_Y$ the leptons transform as  follows

\beqn
\psi_L\equiv \left(
\begin{array}{c}
 \nu_L\\
 \tau_L
\end{array}\right) \sim(1,2,- \frac{1}{2}), \tau^c_L\sim (1,1,1), \nu^c_L\sim (1,1,0),
\eeqn
where the last entry on the right hand side of each $\sim$ is the value of the hypercharge
 $Y$ defined so that $Q=T_3+ Y$. 
These leptons have $V-A$ interactions. Let us now consider mirror
leptons which have $V+A$ interactions. Their quantum numbers are
as follows
\beqn
\chi^c\equiv \left(
\begin{array}{c}
 E_{\tau L}^c\\ 
 N_L^c
\end{array}\right)
\sim(1,2,\frac{1}{2}), E_{\tau L}\sim (1,1,-1), N_L\sim (1,1,0).
\eeqn

The analogous relations for the quarks are 
\beqn
q\equiv \left(
\begin{array}{c}
 t_L\\ 
 b_L
\end{array}\right)
\sim(3,2,\frac{1}{6}), t^c_L\sim (3^*,1,-\frac{2}{3}), b^c_L\sim (3^*,1,\frac{1}{3}),
\eeqn
and for the mirror quarks
\beqn
{Q}^c \equiv \left(
\begin{array}{c}
 B^c_L\\ 
 T^c_L 
\end{array}\right)
\sim(3^*,2,-\frac{1}{6}), T_L\sim (3,1,\frac{2}{3}), B_L\sim (3^*,1, -\frac{1}{3}).
\eeqn
For the Higgs multiplets  we have the MSSM Higgs  doublets which give

\beqn
H_1\equiv 
\br
  H_1^1\\
 H_1^2
 \er
\sim(1,2,-\frac{1}{2}), ~H_2\equiv 
 \br H_2^1\\ 
 H_2^2\er
\sim(1,2,\frac{1}{2}).
\eeqn

We assume that the mirror  generation escapes acquiring mass at the GUT scale and remains
light down to the elctroweak scale where the superpotential of the model
for the lepton part,
 may be written 
in the form
\beqn
W= \epsilon_{ij}  [f_{1} \hat H_1^{i} \hat \psi_L ^{j}\hat \tau^c_L
 +f_{1}' \hat H_2^{j} \hat \psi_L ^{i} \hat \nu^c_L
+f_{2} \hat H_1^{i} \hat \chi^c{^{j}}\hat N_{L}
 +f_{2}' \hat H_2^{j} \hat \chi^c{^{i}} \hat E_{\tau L}]\nonumber\\
+ f_{3} \epsilon_{ij}  \hat \chi^c{^{i}}\hat\psi_L^{j}
 + f_{4} \hat \tau^c_L \hat  E_{\tau L}  +  f_{5} \hat \nu^c_L \hat N_{L}.
\label{superpotential}
\eeqn
In the above we have assumed mixings between the third  generation and the mirror generation.
Such mixings can arise via non-renormalizable interactions\cite{Senjanovic:1984rw}.
Consider, for example,  a term  such as 
$1/M_{Pl} \nu^c_LN_L \Phi_1\Phi_2$. If $\Phi_1$ and $\Phi_2$ develop VEVs of size $10^{9-10}$,
a mixing term of the right size can be generated.

To get the mass matrices of the leptons and the mirror leptons we 
replace the superfields in the superpotential by their component scalar
fields. The relevant parts in the superpotential that produce the lepton and
mirror lepton mass matrices are
\beqn
W=f_1 H_1^1 \tilde{\tau}_L \tilde{\tau}_R^* +f_1' H_2^2 \tilde{\nu}_{L} \tilde{\nu}^*_{R}+
f_2 H_1^1 \tilde{N}_R^* \tilde{N}_L+f_2' H_2^2 \tilde{E}^*_{\tau R} \tilde{E}_{\tau L}\nonumber\\
+f_3 \tilde{E}^*_{\tau R} \tilde{\tau}_L -f_3 \tilde{N}_R^* \tilde{\nu}_{L}+ f_4 \tilde{\tau}_R^* \tilde{E}_{\tau L}
+f_5 \tilde{\nu}^*_{R} \tilde{N}_L
\eeqn
The mass terms for the lepton and their mirrors arise from the part of the lagrangian
\beq
{\cal{L}}=-\frac{1}{2}\frac{\partial ^2 W}{\partial{A_i}\partial{A_j}}\psi_ i \psi_ j+H.c.
\eeq
where $\psi$ and $A$ stand for generic two-component fermion and scalar fields.
After spontaneous breaking of the electroweak symmetry, ($<H_1^1>=v_1/\sqrt{2} $ and $<H_2^2>=v_2/\sqrt{2}$),
we have the following set of mass terms written in 4-spinors for the fermionic sector

\beqn
-{\cal L}_m = \br\bar \tau_R ~ \bar E_{\tau R} \er
 \br
  f_1 v_1/\sqrt{2} ~ f_4\\
 f_3 ~ f_2' v_2/\sqrt{2}\er
 \br \tau_L\\
 E_{\tau L}\er
  + \br\bar \nu_R ~ \bar N_R\er
 \br f'_1 v_2/\sqrt{2} ~ f_5\\
 -f_3 ~ f_2 v_1/\sqrt{2}\er \br \nu_L\\
 N_L\er  + H.c.
\eeqn

Here 
the mass matrices are not  Hermitian and one needs
to use bi-unitrary transformations to diagonalize them. Thus we write the linear transformations

\beqn
 \br\tau_R\\ 
 E_{\tau R}\er=D^{\tau}_R \br\tau_{1_R}\\
 E_{\tau 2_R} \er,\nonumber\\
\br \tau_L\\
 E_{\tau L}\er=D^{\tau}_L \br \tau_{1_L}\\
 E_{\tau 2_L}\er,
\eeqn

such that
\beq
D^{\tau \dagger}_R \br f_1 v_1/\sqrt{2} ~ f_4\\
 f_3 ~ f_2' v_2/\sqrt{2}\er D^{\tau}_L=diag(m_{\tau_1},m_{\tau_2}).
\label{put1}
\eeq

The same holds for the neutrino mass matrix 
\beq
D^{\nu \dagger}_R \br f'_1 v_2/\sqrt{2} ~ f_5\\
 -f_3 ~ f_2v_1/\sqrt{2}\er D^{\nu}_L=diag(m_{\nu_1},m_{\nu_2}).
\label{put2}
\eeq

Here $\tau_1, \tau_2$ are the mass eigenstates and we identify the tau lepton 
with the eigenstate 1, i.e.,  $\tau=\tau_1$, and identify $\tau_2$ with a heavy 
mirror eigenstate  with a mass in the hundreds  of GeV. Similarly 
$\nu_1, \nu_2$ are the mass eigenstates for the neutrinos, 
where we identify $\nu_1$ with the light neutrino state and $\nu_2$ with the 
heavier mass eigen state.
By multiplying Eq.(\ref{put1}) by $D^{\tau \dagger}_L$ from the right and by
$D^{\tau}_R$ from the left and by multiplying Eq.(\ref{put2}) by $D^{\nu \dagger}_L$
from the right and by $D^{\nu}_R$ from the left, one can equate the values of the parameter
$f_3$ in both equations and we can get the following relation
between the diagonlizing matrices $D^{\tau}$ and $D^{\nu}$
\beq
m_{\tau 1} D^{\tau}_{R 21} D^{\tau *}_{L 11} +m_{\tau 2} D^{\tau}_{R 22} D^{\tau *}_{L 12}=
-[m_{\nu 1} D^{\nu}_{R 21} D^{\nu *}_{L 11} +m_{\nu 2} D^{\nu}_{R 22} D^{\nu *}_{L 12}].
\label{condition}
\eeq
 Eq.(\ref{condition}) is an important relation as it constraints the symmetry breaking parameters
 and this constraint must be taken into account in numerical analyses.
 
Let us now write the charged current interaction in the leptonic sector for the 3rd generation 
and for the mirror generation with the W boson. 
\beqn
{\cal{L}}_{CC}= -\frac{g_2}{2\sqrt 2} W^{\dagger}_{\mu}
\left[ \bar \nu\gamma^{\mu} (1-\gamma_5) \tau 
+\bar N\gamma^{\mu} (1+\gamma_5) E_{\tau} \right] + H.c.
\eeqn
In the mass diagonal basis the charged current interactions are given by 
\beqn
{\cal{L}}_{CC}=-\frac{g_2}{2\sqrt 2} W^{\dagger}_{\mu}
\sum_{\alpha,\beta,\gamma,\delta=1,2} \bar \nu_{\alpha} \gamma^{\mu} 
[D^{\nu\dagger}_{L\alpha\gamma} g_{\gamma\delta}^L D^{\tau}_{L\delta\beta} (1-\gamma_5)+ \nonumber\\
+ D^{\nu\dagger}_{R\alpha\gamma} g_{\gamma\delta}^R D^{\tau}_{R\delta\beta} (1+\gamma_5)] 
 \tau_{\beta} +H.c. 
\label{LR}
\eeqn
where $g^{L,R}_{\alpha\beta}$ are defined so that 
\beqn
g^L_{11}=1, g^L_{12}=0= g^L_{21}=g^L_{22}, \nonumber\\
g^R_{11}=0= g^R_{12}= g^R_{21}, g^R_{22}=1.
\eeqn

Next we consider the chargino interactions of the mirror leptons. 
The interaction terms in two-component notation is
\beq
{\cal{L}}=ig\sqrt{2} T^a_{ij} \lambda^a \psi_j A^*_i-\frac{1}{2}\frac{\partial ^2 W}{\partial{A_i}\partial{A_j}}\psi_ i \psi_ j+H.c.
\label{general}
\eeq
Here $T^a=\tau^a/2$ where $\tau^a$ (a=1,2,3) are the Pauli matrices,  and for
 the chargino interaction we use the generators $T^1$ and $T^2$, and 
 $W$ is the part of Eq.(\ref{superpotential}) given by
\beq
W=
-f_2 H_1^2  \tilde{E}^*_{\tau R} \tilde{N}_L -f_2' H_2^1 \tilde{N}^*_R \tilde{E}_{\tau L}.
\eeq
Using the above superpotential and the fermions of the mirror generation
and the supersymmetric partners of the charged Higgs
 for $\psi$ and the mirror sleptons and charged Higgs for $A$,
the interaction of the $V+A$ fourth generation with charginos in the two-component notation  is given by
\beqn
{\cal{L}}=ig[\lambda^+ N^c_L \tilde{E}_{\tau R} +\lambda^- E^c_{\tau L} \tilde{N}_R]\nonumber\\
+\frac{gm_{N}}{\sqrt{2}M_W\cos\beta}[\tilde{N}_L \psi_{H_1^-} E^c_{\tau L}+\tilde{E}^*_{\tau R} \psi_{H_1^-}N_L]\nonumber\\
+\frac{gm_{E}}{\sqrt{2}M_W\sin\beta}[\tilde{N}^*_R \psi_{H_2^+} E_{\tau L}+\tilde{E}_L \psi_{H_2^+}N^c_L]+H.c.,
\label{fourth1}
\eeqn
where $\lambda^{\pm}=\frac{\lambda^1\mp i\lambda^2}{\sqrt{2}}$.

Now we go from two-spinor to four-spinor by defining the two four-spinors:

\beqn
\tilde{W}= 
 \br -i\lambda^+\\
 i\bar \lambda^-\er
, ~\tilde{H}= 
 \br\psi_{H_2^+}\cr 
 \bar \psi_{H_1^-}\er.
\eeqn

By using these two four-spinors,  Eq. (\ref{fourth1}) for the $V+A$ generation interaction 
is given by
\beqn
{\cal{L}}=-g[\bar{\tilde{W}} P_R N \tilde{E}^*_{\tau R}+\bar{\tilde{W^c}} P_R E_{\tau} \tilde{N}^*_R]\nonumber\\
+\frac{gm_{E}}{\sqrt{2} M_W \sin\beta}
[\bar{\tilde{H}} P_R N \tilde{E}^*_{\tau L}+\bar{E_{\tau}} P_R \tilde{H^c} \tilde{N}_R]\nonumber\\
+\frac{gm_{N}}{\sqrt{2} M_W \cos\beta}
[\bar{N} P_R \tilde{H} \tilde{E}_{\tau R}+\bar{\tilde{H^c}} P_R E_{\tau} \tilde{N}^*_L]+H.c.
\label{fourth21}
\eeqn
Now we use the two-component mass eigen states
\beqn
\psi^+_1=-i\lambda^+,~\psi^+_2=\psi_{H^+_2}\nonumber\\
\psi^-_1=-i\lambda^-,~\psi^-_2=\psi_{H^-_1}
\eeqn
By defining the two-component spinors $\chi_i^+$ and $\chi_i^-$ as
\beqn
\chi^+_i=V_{ij}\psi^+_j\nonumber\\
\chi^-_i=U_{ij}\psi^-_j
\eeqn
the four-component mass eigen states are

\beqn
\tilde{\chi_1}^+= 
 \br\chi^+_1\\
 \bar \chi^-_1\er
, ~\tilde{\chi_2}^+= 
 \br\chi^+_2\cr 
 \bar\chi^-_2\er
\eeqn

The matrix elements $U$ and $V$ that  diagonalize the chargino mass matrix $M_C$ are
given by
\beq
U^* M_C V^{-1}= diag (m_{\tilde{\chi_1}}^+,m_{\tilde{\chi_2}}^+).
\eeq
One can use the definitions of $P_L$, $P_R$ and the above relations to get the following useful
relations
\beqn
P_L \tilde{W}=P_L \sum_{i=1}^2 V^*_{i1}\tilde{\chi_i}^+,~
P_L \tilde{W}^c=P_L\sum_{i=1}^2 U^*_{i1}\tilde{\chi_i}^c \nonumber\\
P_L \tilde{H}=P_L \sum_{i=1}^2V^*_{i2}\tilde{\chi_i}^,~
P_R \tilde{H}=P_R \sum_{i=1}^2 U_{i2}\tilde{\chi_i}^+\nonumber\\
P_R \tilde{H}^c=P_R \sum_{i=1}^2 V_{i2}\tilde{\chi_i}^c,~
P_L \tilde{H}^c=P_L \sum_{i=1}^2 U^*_{i2}\tilde{\chi_i}^c
\eeqn

Using these relations and Eq.(\ref{fourth21}),
the interactions of the mirror generation with chargino mass-eigen states is given by 
\beqn
-{\cal{L}}_{N - E_{\tau}- \chi^+}= g\bar N [V^*_{i1} P_L -\kappa_{N} U_{i2} P_R] \tilde{\chi_i}^+ \tilde E_{\tau R}\nonumber\\ 
+ g\bar N [ -\kappa_{E_{\tau}} V^*_{i2} P_L] \tilde{\chi_i}^+ \tilde E_{\tau L}
+ g\bar E_{\tau} [U^*_{i1} P_L -\kappa_{E_{\tau}} V_{i2}  P_R] \tilde{\chi_i}^c \tilde N_R\nonumber\\ 
+ g\bar E_{\tau} [ -\kappa_{N} U^*_{i2} P_L] \tilde{\chi_i}^c \tilde N_L
+ H.c.
\label{c2}
\eeqn
where $\tilde{\chi_i}^c$ is the charge conjugate of $\tilde {\chi_i}$ and where
\beqn
\kappa_N=\frac{m_N}{\sqrt{2} M_W \cos\beta},~
\kappa_{E_{\tau}}=\frac{m_{E_{\tau}}}{\sqrt{2} M_W \sin\beta}
\eeqn

The interaction of the leptons with the chargino is given by
\beqn
-{\cal{L}}_{\nu - \tau- \chi^+}= g\bar \nu [U_{i1} P_R -\kappa_{\nu} V_{i2}^*  P_L] \tilde{\chi_i}^+ \tilde \tau_L\nonumber\\ 
+ g\bar \nu [ -\kappa_{\tau} U_{i2} P_R] \tilde{\chi_i}^+ \tilde \tau_R
+ g\bar \tau [V_{i1} P_R -\kappa_{\tau} U_{i2}^*  P_L] \tilde{\chi_i}^c \tilde \nu_L\nonumber\\ 
+ g\bar \tau [ -\kappa_{\nu} V_{i2} P_R] \tilde{\chi_i}^c \tilde \nu_R  
+ H.c.,
\label{c1}
\eeqn
where
\beqn
\kappa_{\tau}=\frac{m_{\tau}}{\sqrt{2} M_W \cos\beta},~\kappa_{\nu}=
\frac{m_{{\nu}}}{\sqrt{2} M_W \sin\beta}.
\eeqn
  A full analysis of the mirror sparticle couplings will be given elsewhere.

Next we  consider  the mixings of the charged sleptons and the charged mirror sleptons. 
The mass matrix in the basis $(\tilde  \tau_L, \tilde E_L, \tilde \tau_R, 
\tilde E_R)$ takes the form 

\beqn
(M^2)_{\tilde \tau}= \br M^2_{11} ~ M^2_{12} ~ M^2_{13} ~ M^2_{14} \\
M^{2}_{21} ~ M^2_{22} ~ M^2_{23} ~ M^2_{24} \\ 
M^{2}_{31}  ~ M^2_{32} ~ M^2_{33} ~ M^2_{34}\\
M^2_{41} ~ M^2_{42} ~M^2_{43} ~ M^2_{44}
\er.
\eeqn
Here the terms $M^2_{11}, M^2_{13}, M^2_{31}, M^2_{33}$ arise from soft  
breaking in the  sector $\tilde \tau_L, \tilde \tau_R$. Similarly the terms 
$M^2_{22}, M^2_{24},$  $M^2_{42}, M^2_{44}$ arise from soft  
breaking in the  sector $\tilde E_L, \tilde E_R$. The terms $M^2_{12}, M^2_{21},$
$M^2_{23}, M^2_{32}$, $M^2_{14}, M^2_{41}$, $M^2_{34}, M^2_{43},$  arise  from mixing between the staus  and 
the mirrors.  We assume that all the masses are of the electroweak scale 
so all the terms enter in the diagonalization.  We diagonalize the hermitian mass$^2$ matrix  by the
following unitary transformation 
\beqn
 \tilde D^{\tau \dagger} M^2_{\tilde \tau} \tilde D^{\tau} = diag (M^2_{\tilde \tau_1},  
M^2_{\tilde \tau_2}, M^2_{\tilde \tau_3},  M^2_{\tilde \tau_4}).
\eeqn
A similar mass matrix exists in the sneutrino sector.
In the basis $(\tilde  \nu_L, \tilde N_L, \tilde \nu_R, 
\tilde N_R)$ it takes the form 

\beqn
(M^2)_{\tilde \nu}= \br m^2_{11} ~ m^2_{12} ~ m^2_{13} ~ m^2_{14} \\
m^{2}_{21} ~ m^2_{22} ~ m^2_{23} ~ m^2_{24} \\
m^{2}_{31} ~ m^2_{32} ~ m^2_{33} ~ m^2_{34}\\ 
m^2_{41} ~ m^2_{42} ~ m^2_{43} ~ m^2_{44}
\er.
\eeqn
As in the charged  slepton sector 
here also the terms $m^2_{11}, m^2_{13}, m^2_{31}, m^2_{33}$ arise from soft  
breaking in the  sector $\tilde \nu_L, \tilde \nu_R$. Similarly the terms 
$m^2_{22}, m^2_{24},$  $m^2_{42}, m^2_{44}$ arise from soft  
breaking in the  sector $\tilde N_L, \tilde N_R$. The terms $m^2_{12}, m^2_{21},$
$m^2_{23}, m^2_{32}$, $m^2_{14}, m^2_{41}$, $m^2_{34}, m^2_{43},$  arise  
from mixing between the physical sector and 
the mirror sector.  Again as in the charged lepton sector 
we assume that all the masses are of the electroweak size
so all the terms enter in the diagonalization.  The above matrix can be diagonalized  by the
following unitary transformation 
\beqn
 \tilde D^{\nu\dagger} M^2_{\tilde \nu} \tilde D^{\nu} = diag (M^2_{\tilde \nu_1},  
M^2_{\tilde \nu_2}, M^2_{\tilde \nu_3},  M^2_{\tilde \nu_4}).
\eeqn
The physical tau and neutrino states are $\tau\equiv \tau_1, \nu\equiv \nu_1$,
and the states $\tau_2, \nu_2$ are heavy states with mostly mirror particle content. 
The states $\tilde \tau_i, \tilde \nu_i; ~i=1-4$ are the slepton and sneutrino states. 
For the case of  
no mixing these limit as  follows 
\beqn
\tilde \tau_1\to \tilde \tau_L, ~\tilde \tau_2\to \tilde E_L, ~\tilde \tau_3\to \tilde \tau_R, ~
\tilde \tau_4\to \tilde E_R\nonumber\\
\tilde \nu_1\to \tilde \nu_L, ~\tilde \nu_2\to \tilde N_L, ~\tilde \nu_3\to \tilde \nu_R, ~
\tilde \nu_4\to \tilde N_R.
\eeqn
A further 
 discussion of the scalar mass$^2$ matrices is given in Appendix A.

In the  mass diagonal basis the interactions of the neutrino $\nu$  and of the 
stau which include
the mixing effects with the mirrors are given by
\beqn
-{\cal{L}}_{\nu - \tilde{\tau}- \chi^+}= 
\sum_{\alpha =1-2} \sum_{j=1-4}  
g\bar\nu_{\alpha}[
D^{\nu\dagger}_{L \alpha 1}U_{i1}P_R- D^{\nu\dagger}_{R \alpha 1}\kappa_{\nu}V^*_{i2}P_L]\tilde{\chi}^+_i
\tilde D^{\tau}_{1 j}\tilde \tau_j\nonumber\\
+g\bar\nu_{\alpha}[-
 D^{\nu\dagger}_{L \alpha 1}\kappa_{\tau}U_{i2}P_R]\tilde{\chi}^+_i
\tilde D^{\tau}_{3 j}\tilde \tau_j\nonumber\\
+g\bar\nu_{\alpha}[
D^{\nu\dagger}_{R \alpha 2}V^*_{i1}P_L- D^{\nu\dagger}_{L \alpha 2}\kappa_{N}U_{i2}P_R]\tilde{\chi}^+_i
\tilde D^{\tau}_{4 j}\tilde \tau_j\nonumber\\
+g\bar\nu_{\alpha}[-
 D^{\nu\dagger}_{R \alpha 2}\kappa_{E_{\tau}}V^*_{i2}P_L]\tilde{\chi}^+_i
\tilde D^{\tau}_{2 j}\tilde \tau_j+H.c
 \label{chargino}
  \eeqn
For ${\cal{L}}_{\tau - \tilde{\nu}- \chi^+}$ we have
\beqn
-{\cal{L}}_{\tau - \tilde{\nu}- \chi^+}= 
\sum_{\alpha =1-2} \sum_{j=1-4}
g\bar\tau_{\alpha}[
D^{\tau\dagger}_{L \alpha 1}V_{i1}P_R- D^{\tau\dagger}_{R \alpha 1}\kappa_{\tau}U^*_{i2}P_L]\tilde{\chi}^c_i
\tilde D^{\nu}_{1 j}\tilde \nu_j\nonumber\\
+g\bar\tau_{\alpha}[-
 D^{\tau\dagger}_{L \alpha 1}\kappa_{\nu}V_{i2}P_R]\tilde{\chi}^c_i
\tilde D^{\nu}_{3 j}\tilde \nu_j\nonumber\\
+g\bar\tau_{\alpha}[
D^{\tau\dagger}_{R \alpha 2}U^*_{i1}P_L- D^{\tau\dagger}_{R \alpha 2}\kappa_{E_{\tau}}V_{i2}P_R]\tilde{\chi}^c_i
\tilde D^{\nu}_{4 j}\tilde \nu_j\nonumber\\
+g\bar\tau_{\alpha}[-
 D^{\tau\dagger}_{R \alpha 2}\kappa_{N}U^*_{i2}P_L]\tilde{\chi}^c_i
\tilde D^{\nu}_{2 j}\tilde \nu_j+H.c
\label{charginoa}
  \eeqn

Next we look at the neutral current interactions and focus on the charged
	leptons. Here the Z  boson interactions are given by

	\beqn
	{\cal{L}}_{NC}= -\frac{g}{4 \cos\theta_W} Z_{\mu}\left[ \bar \tau\gamma^{\mu} (4x-1+\gamma_5)\tau
	+  \bar E_{\tau}\gamma^{\mu} (4x-1-\gamma_5)E_{\tau} \right],
	\label{nc}
	\eeqn
where $x=\sin^2\theta_W$.  We write the result in the mass diagonal 
basis and get

\beqn
{\cal{L}}_{NC}=- \frac{g}{2 \cos\theta_W} Z_{\mu}
\sum_{\alpha=1,2}\sum_{\beta=1,2} ( \bar \tau_{\alpha}\gamma^{\mu} \tau_{\beta}) \nonumber\\
(x\{D^{\tau \dagger}_{L \alpha 1}D^{\tau}_{L 1\beta}+D^{\tau \dagger}_{R \alpha 1}D^{\tau}_{R 1\beta}
+D^{\tau \dagger}_{L \alpha 2}D^{\tau}_{L 2\beta}+D^{\tau \dagger}_{R \alpha 2}D^{\tau}_{R 2\beta}
\}\nonumber\\
-\frac{1}{2}\{D^{\tau \dagger}_{L \alpha 1}D^{\tau}_{L 1\beta}+D^{\tau \dagger}_{R \alpha 2}D^{\tau}_{R 2\beta}\})\nonumber\\
+( \bar \tau_{\alpha}\gamma^{\mu}\gamma_5 \tau_{\beta})
(x\{-D^{\tau \dagger}_{L \alpha 1}D^{\tau}_{L 1\beta}+D^{\tau \dagger}_{R \alpha 1}D^{\tau}_{R 1\beta}
-D^{\tau \dagger}_{L \alpha 2}D^{\tau}_{L 2\beta}+D^{\tau \dagger}_{R \alpha 2}D^{\tau}_{R 2\beta}\}\nonumber\\
+\frac{1}{2}\{D^{\tau \dagger}_{L \alpha 1}D^{\tau}_{L 1\beta}-D^{\tau \dagger}_{R \alpha 2}D^{\tau}_{R 2\beta}\}).
\label{zinteractions}
\eeqn

Next we discuss the neutralino interaction. Using the parts of Eq. (\ref{general})
that produce the interaction of the mirror lepton with the neutralino we have
\beq
{\cal{L}}=i\frac{g}{\sqrt{2}} \tau^3_{ij} \lambda^3 \psi_j A^*_i
+ig'\sqrt{2}Y_i\delta_{ij}\lambda'\psi_j A^*_i
-\frac{1}{2}\frac{\partial ^2 W}{\partial{A_i}\partial{A_j}}\psi_ i \psi_ j+H.c.
\eeq 
The part of interest in the superpotential here is
\beqn
W=
f_2 H_1^1 \tilde{N}_R^* \tilde{N}_L+f_2' H_2^2 \tilde{E}^*_{\tau R} \tilde{E}_{\tau L}
\eeqn
By using the fermions of the mirror generation and the supersymmetric partners of the neutral Higgs for $\psi$
 and the mirror sleptons and neutral Higgs for $A$
  one gets the following
lagrangian for
the interactions of the mirror leptons with neutralino in the two component notation
\beqn
{\cal{L}}=i\frac{g}{\sqrt{2}}\lambda^3
[E^c_{\tau L} \tilde{E}_{\tau R}-N^c_L\tilde{N}_R]
+i\frac{g'}{\sqrt{2}}\lambda'
[E^c_{\tau L} \tilde{E}_{\tau R}+N^c_L\tilde{N}_R]\nonumber\\
-i\sqrt{2}g'\lambda' E_{\tau L} \tilde{E}^*_{\tau L}
-\frac{gm_{N}}{\sqrt{2}M_W\cos\beta}[\tilde{N}_L \psi_{H_1^0} N^c_{L}+\tilde{N}^*_{R} \psi_{H_1^0}N_L]\nonumber\\
-\frac{gm_{E}}{\sqrt{2}M_W\sin\beta}[\tilde{E}_{\tau L} \psi_{H_2^0} E^c_{\tau L}+\tilde{E}^*_{\tau R} \psi_{H_2^0}\tilde{E}_{\tau L}]+H.c.
\label{fourth2}
\eeqn
Now we go from two-spinor to four-spinor  by defining the four Majorana spinors
\beqn
\tilde{B}= 
 \br-i\lambda'\\
 i\bar \lambda'\er
, ~\tilde{W}_3= 
\br-i\lambda^3\cr 
 i\bar \lambda^3\er
,~\tilde{H}_1=
 \br\psi_{H_1^0}\cr 
 \bar \psi_{H_1^0}\er
,~\tilde{H}_2=
 \br \psi_{H_2^0}\cr 
 \bar \psi_{H_2^0}\er.
\eeqn

The lagrangian in terms of these fields reads
\beqn
{\cal{L}}=\frac{1}{\sqrt{2}}\tilde{N}_R [g\bar N P_L \tilde{W}_3-g'\bar N P_L \tilde{B}]
-\frac{1}{\sqrt{2}}\tilde{E}_{\tau R} [g\bar E_{\tau} P_L \tilde{W}_3+g'\bar E_{\tau} P_L \tilde{B}]\nonumber\\
+\sqrt{2}g' \tilde{E}^*_{\tau L} \bar{\tilde{B}} P_L E_{\tau}
-\frac{gm_{N}}{\sqrt{2}M_W\cos\beta}[\tilde{N}_L \bar N P_L \tilde{H_1}+
\tilde{N}^*_R \bar{\tilde{H_1}} P_L N]\nonumber\\
-\frac{gm_{E}}{\sqrt{2}M_W\sin\beta}[\tilde{E}_{\tau L} \bar{E_{\tau}} P_L \tilde{H_2}+
\tilde{E}^*_{\tau R} \bar{\tilde{H_2}} P_L E_{\tau}].
\label{neutralino2}
\eeqn
We can write this interaction in the neutralino mass eigen state basis $\tilde{\chi}^0_j$ where
\beq
X^T M_{\tilde{\chi}^0} X=diag(m_{{\chi^0}_1}, m_{{\chi^0}_2}, m_{{\chi^0}_3}, m_{{\chi^0}_4})
\eeq
In  writing Eq.(\ref{neutralino2}) in this basis the following relations are found  useful
\beqn
P_L \tilde{W}_3 = P_L \sum_{j=1}^4 X_{2j} \tilde{\chi}^0_j,~P_L \tilde{B} = P_L \sum_{j=1}^4 X_{1j} \tilde{\chi}^0_j,\nonumber\\
P_L \tilde{H}_1 = P_L \sum_{j=1}^4 X_{3j} \tilde{\chi}^0_j,
P_L \tilde{H}_2 = P_L \sum_{j=1}^4 X_{4j} \tilde{\chi}^0_j,\nonumber\\
\bar{\tilde{H_1}} P_L=\sum_{j=1}^4 X_{3j}\bar{\tilde{\chi}}^0_j P_L,~
\bar{\tilde{H_2}} P_L=\sum_{j=1}^4 X_{4j}\bar{\tilde{\chi}}^0_j P_L,~
\bar{\tilde{B}} P_L=\sum_{j=1}^4 X_{1j}\bar{\tilde{\chi}}^0_j P_L
\eeqn 
Using the above 
the interactions of the mirror lepton $E_{\tau}$ with the neutralino mass eigen states is given by
\beqn
-{\cal{L}}_{E_{\tau}-\tilde{E}_{\tau}-\chi^0}=
\frac{1}{\sqrt 2} \sum_{j=1-4}\left[
\bar E_{\tau}\left(a'_j- b'_j\gamma_5\right)  \tilde  \chi^0_j \tilde E_{\tau R} 
+ \bar E_{\tau}\left(c'_j- d'_j\gamma_5\right)  \tilde  \chi^0_j \tilde  E_{\tau L}
\right]+H.c.
\eeqn
Here 
\beqn
a_j'= (\alpha_{E_{\tau}j} + \beta_{E_{\tau}j}),~~ 
b_j'=(-\alpha_{E_{\tau}j} + \beta_{E_{\tau}j}), \nonumber\\
c_j'=-(\gamma_{E_{\tau}j} +\delta_{E_{\tau}j}), 
d_j'= (\gamma_{E_{\tau}j} -\delta_{E_{\tau}j}),
\eeqn
and   $\alpha_{E_{\tau j}}$, $\beta_{E_{\tau j}}$, $\gamma_{E_{tau j}}$ and $\delta_{E_{\tau j}}$ are 
defined so that 
\beqn\label{alphabk}
\alpha_{E_{\tau j}} =\frac{g m_{E} X^*_{4j}}{2m_W\sin\beta},~~
\beta_{E_{\tau j}}=eX_{1j}^{'} +\frac{g}{\cos\theta_W} X_{2j}^{'}
(\frac{1}{2}-\sin^2\theta_W),\nonumber\\
\gamma_{E_{\tau j}}=e X_{1j}^{'*}-\frac{g\sin^2\theta_W}{\cos\theta_W}
X_{2j}^{*'},
~~ \delta_{E_{\tau j}}=-\frac{g m_{E} X_{4j}}{2m_W \sin\beta}
\eeqn
and
\beqn
X'_{1j}= (X_{1j}\cos\theta_W + X_{2j} \sin\theta_W), \nonumber\\
X'_{2j}=  (-X_{1j}\sin\theta_W + X_{2j} \cos\theta_W).
\eeqn
The above may be compared with the 
 interactions of the  $\tau$ lepton with  neutralinos which are given by
\beqn
-{\cal{L}}_{\tau-\tilde \tau-\chi^0}= \frac{1}{\sqrt 2} \sum_{j=1-4}\left[
\bar \tau\left(a_j+ b_j\gamma_5\right)  \tilde  \chi^0_j \tilde \tau_L 
+ \bar \tau\left(c_j+ d_j\gamma_5\right)  \tilde  \chi^0_j \tilde  \tau_R
\right]+H.c
\eeqn
Here
\beqn
a_j= (\alpha_{\tau j} + \beta_{\tau j}), ~~b_j=(-\alpha_{\tau j} +\beta_{\tau j}), \nonumber\\
c_j=-(\gamma_{\tau j} +\delta_{\tau j}), ~~d_j= (\gamma_{\tau j}- \delta_{\tau j}),
\eeqn
where
\beqn
\alpha_{\tau j} =\frac{g m_{\tau} X_{3j}}{2m_W\cos\beta},~~
\beta_{\tau j}=-eX_{1j}^{'*} +\frac{g}{\cos\theta_W} X_{2j}^{'*}
(-\frac{1}{2}+\sin^2\theta_W),\nonumber\\
\gamma_{\tau j}=-e X_{1j}'+\frac{g\sin^2\theta_W}{\cos\theta_W}
X_{2j}',
~~ \delta_{\tau j}=-\frac{g m_{\tau} X^*_{3j}}{2m_W \cos\beta}.
\eeqn
Rotation into the mass diagonal basis of the leptons and sleptons gives the result
\beqn
-{\cal{L}}_{\tau-\tilde \tau-\chi^0} = \frac{1}{\sqrt 2} 
\sum_{\alpha =1-2} 
\sum_{k=1-4} \sum_{j=1-4}
\bar \tau_{\alpha}   [(D^{\tau\dagger}_+)_{\alpha 1} a_j 
 + (D^{\tau\dagger}_-)_{\alpha 1} b_j \nonumber\\
+\gamma_5 \left((D^{\tau\dagger}_-)_{\alpha 1} a_j + (D^{\tau\dagger}_+)_{\alpha 1} b_j \right)]
 \tilde  \chi^0_j  (\tilde D^{\tau})_{1 k} \tilde \tau_k\nonumber\\
+\bar \tau_{\alpha}   [(D^{\tau\dagger}_+)_{\alpha 1} c_j + (D^{\tau\dagger}_-)_{\alpha 1} d_j
 +\gamma_5 \left((D^{\tau\dagger}_-)_{\alpha 1} c_j + (D^{\tau\dagger}_+)_{\alpha 1} d_j \right)]
 \tilde  \chi^0_j  (\tilde D^{\tau})_{3 k} \tilde \tau_k\nonumber\\ 
+\bar \tau_{\alpha}   [(D^{\tau\dagger}_+)_{\alpha 2} a'_j - (D^{\tau\dagger}_-)_{\alpha 2} b'_j
 +\gamma_5 \left((D^{\tau\dagger}_-)_{\alpha 2} a'_j - (D^{\tau\dagger}_+)_{\alpha 2} b'_j \right)]
 \tilde  \chi^0_j  (\tilde D^{\tau})_{4 k} \tilde \tau_k\nonumber\\
+\bar \tau_{\alpha}   [(D^{\tau\dagger}_+)_{\alpha 2} c'_j
 - (D^{\tau\dagger}_-)_{\alpha 2} d'_j
 +\gamma_5 \left((D^{\tau\dagger}_-)_{\alpha 2} c'_j - (D^{\tau\dagger}_+)_{\alpha 2} d'_j \right)]
 \tilde  \chi^0_j  (\tilde D^{\tau})_{2 k} \tilde \tau_k +H.c.
\label{neutralinoo}
\eeqn
 where 
 \beqn
 D_{\pm}^{\tau}= \frac{1}{2} (D^{\tau}_L\pm D^{\tau}_R).
 \eeqn
Our final result including the mixings of leptons and mirror leptons and the mixings
of sleptons and of mirror sleptons are given by Eq.(\ref{LR}) for the W boson interactions,  
Eq.(\ref{chargino}) and Eq. (\ref{charginoa}) for the 
chargino interactions  and by Eq.(\ref{zinteractions}) for the Z boson interactions, and 
by Eq.(\ref{neutralinoo}) for the neutralino interactions.

\section{Neutrino magnetic moment\label{magnetic}} 
\begin{figure}
 \vspace{-7cm}
      \scalebox{2.25}
      {
       \hspace{-2cm}      
       \includegraphics[width=10cm,height=12cm]{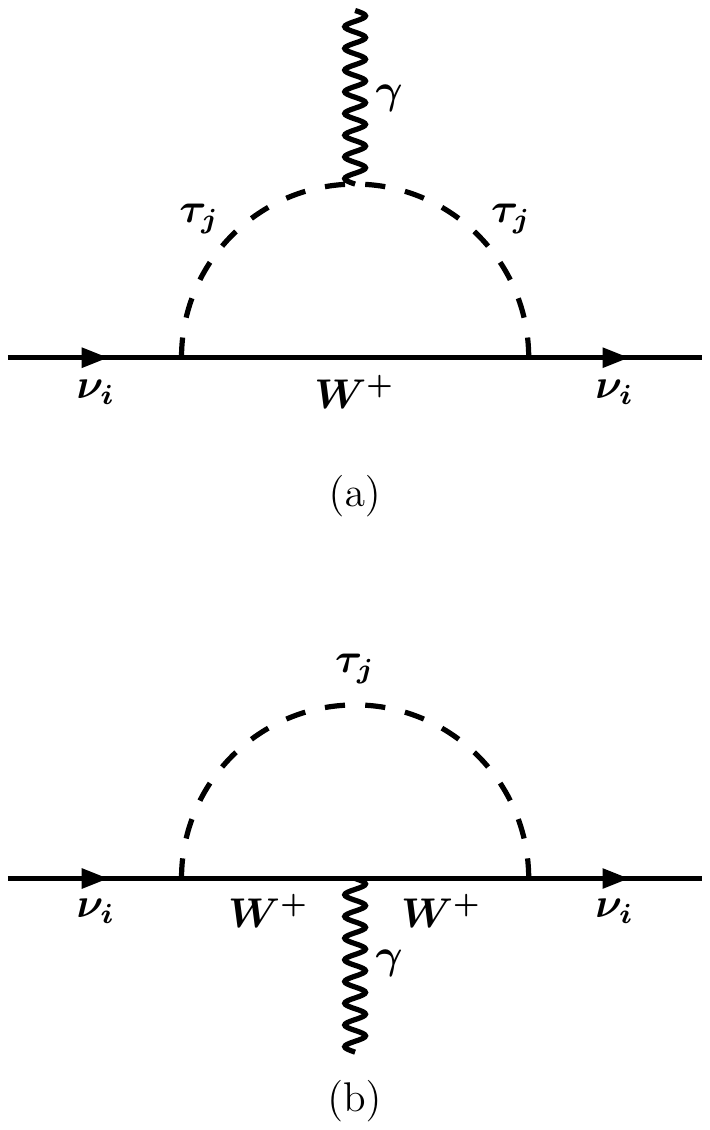}
      }     
      \vspace{-12cm}
\caption{The loop contributions to the magnetic 
dipole moment of neutrinos ($\nu_i$)  via exchange of $W^+$ boson
and via the exchange of leptons and mirror leptons denoted by $\tau_j$.}
\label{smfig}
       \end{figure}

\begin{figure}
 \vspace{-7cm}
      \scalebox{2.25}
      {
       \hspace{-2cm}      
       \includegraphics[width=10cm,height=12cm]{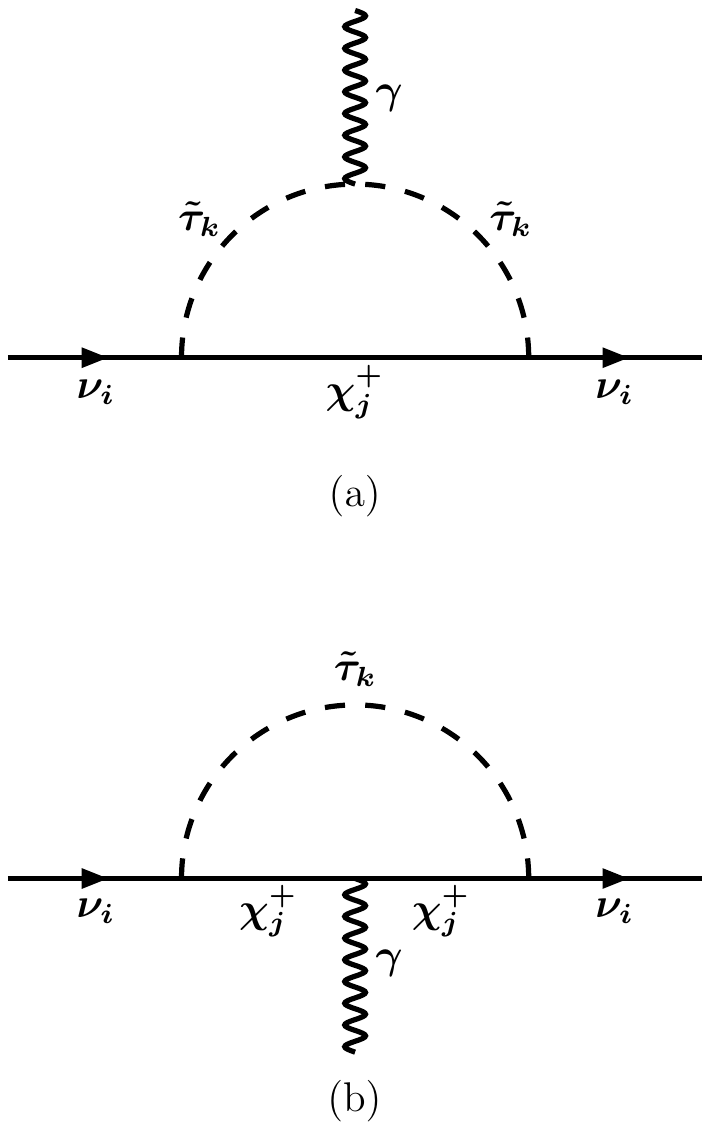}
     }     
      \vspace{-12cm}
\caption{The supersymmetric  loop contributions to the magnetic 
dipole moment of neutrinos ($\nu_i$)  via exchange of charginos ($\chi_j^+$), sleptons and mirror sleptons denoted by $\tilde \tau_k$.}
\label{susyfig}
       \end{figure}

The discovery of neutrino masses from the solar and atmospheric data 
\cite{Abdurashitov:1999zd,Ahmad:2002jz,Altmann:2000ft,Ambrosio:2001je,Fukuda:2000np,Hampel:1998xg}
has very significantly advanced our understanding of the basic nature of these particles. 
One outcome of non-vanishing neutrino masses is the possibility that they could possess
non-vanishing magnetic and electric dipole moments if the neutrinos are Dirac particles 
while only transition magnetic moments are allowed if they are Majorana. In this 
analysis we assume the Dirac nature of the neutrinos. In this case the neutrinos 
will have non-vanishing magnetic and electric dipole moments and 
such moments could
 enter  in several physical   phenomena\cite{pheno}.
One phenomena where the moments may play a role is in the 
neutrino spin flip  processes  such as\cite{Kuznetsov:2007ct}
$\nu_L\to \nu_R +\gamma^*$ or $\nu_L +\gamma^*\to \nu_R$.  
From experiment, there  already  exist  limits on both the magnetic and
the electric dipole moments of neutrinos.
Our focus will be the magnetic moment of the tau neutrino which is affected
by the mixing effects from the mirror leptons. 
(For previous work on neutrino magnetic moment with mirror effects in a different context
see \cite{Maalampi:1988va})
 The current  limits on the magnetic moment of the $\tau$ neutrino is\cite{exp3}
 \beqn
  |\mu(\nu_{\tau})|\leq 1.3\times 10^{-7} \mu_B 
 \eeqn
 where $\mu_B=(e/2m_e)$ is the Bohr magneton. 
The magnetic moment of the neutrino arises  in the Standard Model
at one loop via the exchange of the W boson assuming one extends the
Standard Model  to include a right handed neutrino
(see  Fig.(\ref{smfig})),
and in the supersymmetric models there are  additional 
contributions arising from the chargino exchange 
contributions (see Fig.(\ref{susyfig})).

  Neutrino masses for the first three generations are very small, i.e., from WMAP data one has
   $\sum_i |m_{\nu_i}|\le (.7-1)$ eV\cite{Spergel:2003cb}.   
   If the neurtinos  are Dirac one would need to explain, how such tiny Dirac masses are generated
 which would typically require fine tunings of $O(10^{-10})$ or more.
 However, unlike the Majorana neutrino case for which there  is  a standard mechanism for the
 generation of small neutrino masses, i.e., see-saw,  there is no standard mechanism for the generation of small 
 Dirac neutrino masses. Indeed this topic continues to a subject of ongoing  research and several 
 recent works can be found in \cite{hung,diracmass}. Here, we 
   do not go into details on this topic  which would take us far afield. 
   Thus in this work we do not make any attempt to deduce the smallness  of the neutrino masses
   but rather assume this is the case.
    With this assumption   we  discuss below the tau neutrino 
	 magnetic moment in the extended MSSM with mirrors
	 for the case when there is mixing with
	 the mirror leptons.  The contributions to be discussed arise 
	  from loops containing (1) lepton (mirror lepton)- W boson and (2) scalar leptons (scalar mirrors)- charginos. 
From Eq.(\ref{LR}) one can calculate the W boson, charged lepton and charged
mirror lepton contributions arising from Fig.(1)   to the
 magnetic moment of the $\tau$
neutrino in $\mu_B$ units to be
\beqn
\mu^{(1)}_{\nu}=\frac{-G_Fm_e}{8\pi^2\sqrt{2}}\sum_{\gamma =1}^{2} \sum_{\delta =1}^{2} \sum_{\beta =1}^{2}
m_{\tau_\beta} G_1(\frac{m_{\tau_\beta}}{M_W})\nonumber\\
(|(D^{\nu \dagger}_L)_{1\gamma}g^L_{\gamma \delta}(D^{\tau}_L)_{\delta \beta}
+(D^{\nu \dagger}_R)_{1\gamma}g^R_{\gamma \delta}(D^{\tau}_R)_{\delta \beta}|^2\nonumber\\
-|(D^{\nu \dagger}_L)_{1\gamma}g^L_{\gamma \delta}(D^{\tau}_L)_{\delta \beta}
-(D^{\nu \dagger}_R)_{1\gamma}g^R_{\gamma \delta}(D^{\tau}_R)_{\delta \beta}|^2)\nonumber\\
+\frac{3G_Fm_{\nu}m_e}{16\pi^2\sqrt{2}}
\sum_{\gamma =1}^{2} \sum_{\delta =1}^{2} \sum_{\beta =1}^{2}G_2(\frac{m_{\tau_\beta}}{M_W})\nonumber\\
(|(D^{\nu \dagger}_L)_{1\gamma}g^L_{\gamma \delta}(D^{\tau}_L)_{\delta \beta}
+(D^{\nu \dagger}_R)_{1\gamma}g^R_{\gamma \delta}(D^{\tau}_R)_{\delta \beta}|^2\nonumber\\
+|(D^{\nu \dagger}_L)_{1\gamma}g^L_{\gamma \delta}(D^{\tau}_L)_{\delta \beta}
-(D^{\nu \dagger}_R)_{1\gamma}g^R_{\gamma \delta}(D^{\tau}_R)_{\delta \beta}|^2),
\label{leptonmirror}
\eeqn
where the form factor functions $G_1(r)$ and $G_2(r)$ are given by
\beqn
G_1(r)=\frac{4-r^2}{1-r^2}+\frac{3r^2}{(1-r^2)^2}\ln(r^2),\nonumber\\
G_2(r)=\frac{2-5r^2+r^4}{(1-r^2)^2}-\frac{2r^4}{(1-r^2)^3}\ln(r^2).
\eeqn
As noted already Eq.(\ref{leptonmirror}) includes  the contributions from the tau
 and from the mirror lepton.
 We parametrize the mixing between $\tau$ and $E_{\tau}$ by the angle $\theta$, where
\beqn
{\left(\begin{array}{c}
\tau\cr
E_\tau \end{array}\right)}=
 {\left(
\begin{array}{cc}
\cos\theta & \sin\theta \cr
             -\sin\theta & \cos\theta
\end{array}\right)}{\left(\begin{array}{c}
\tau_1\cr
\tau_2\end{array}\right)},
\eeqn
and the mixing between $\nu$ and $N$ by the angle $\phi$ where
\beqn
{\left(\begin{array}{c}
\nu\cr
N\end{array}\right)}=
 {\left(
\begin{array}{cc}
\cos\phi & \sin\phi \cr
             -\sin\phi & \cos\phi
 \end{array}\right)}{\left(\begin{array}{c}
\nu_1\cr
\nu_2\end{array}\right)}.
\eeqn
where we take $D^{\tau}_L=D^{\tau}_R$ and $D^{\nu}_L=D^{\nu}_R$ or $\theta_L=\theta_R=\theta$ and $\phi_L=\phi_R=\phi$.  These are simplicity assumptions to get the size of numerical estimates and are  easily
improved with better understanding of mixings with mirror and ordinary leptons.
We identify $\tau_1$ with the physical $\tau$ and $\tau_2$ with the mirror  generation
lepton. When there is no risk of confusion we  will set $\tau_1=\tau$ and $\tau_2=E$, and
similarly for the  $\nu_1$ and $\nu_2$ where we set $\nu_1=\nu_{\tau}$ and $\nu_2=N$. 
Now 
we see that the first term of Eq.(\ref{leptonmirror}) is proportional to the fermion
mass $m_{\tau_{\beta}}$ which could be a lepton or a mirror lepton. 
For the lepton loop $\beta =1$, the first term in Eq.(\ref{leptonmirror})
 is proportional to $[\cos^2(\theta -\phi)-\cos^2(\theta +\phi)]$
and the second term is proportional to $[\cos^2(\theta -\phi)+\cos^2(\theta +\phi)]$.
For the mirror lepton loop $\beta =2$, and the first term 
in Eq.(\ref{leptonmirror})
is proportional to $[\sin^2(\theta -\phi)-\sin^2(\theta +\phi)]$
while 
 the second term in Eq.(\ref{leptonmirror})
  is proportional to $[\sin^2(\theta -\phi)+\sin^2(\theta +\phi)]$.
Thus if the mixing between lepton and mirror leptons exist, the first term for the case of $\beta =2$
 can produce  a large contribution to the neutrino magnetic moment if the mirror lepton mass is in
the region of few hundreds GeV. Also if this mixing is absent,  the 
contribution would come only 
  from the $\tau$-lepton loop. In this case, the first term does not contribute and the second term gives the
  result
\beq
\frac{3m_{\tau}m_{\nu_{\tau}}G_F}{4\sqrt 2\pi^2},
\eeq
 taking into account the limit  $G_2(0)=2$. Thus 
Eq.(\ref{leptonmirror}) gives for the neutrino magnetic moment the value of $3.2\times 10^{-19}(\frac{m_{\nu}}{eV})\mu_B$ and agrees with the previous analyses given in the Standard Model
\cite{CabralRosetti:1999ad,Dvornikov:2003js}. We note that the underlying assumptions of \cite{CabralRosetti:1999ad,Dvornikov:2003js} regarding a small Dirac mass is identical to ours except that
our analysis is more general in that it  includes both supersymmetry and mirror contributions.\\  

Next we compute the supersymmetric contributions to the $\nu_{\tau}$ magnetic moment which include the 
chargino, the slepton and the mirror slepton  contributions which can be calculated using 
Eq.(\ref{chargino}). The result in $\mu_B$ units is
\beqn
\mu^{(2)}_{\nu}=-\frac{g^2m_e}{16\pi^2}\sum_{k =1}^{2} \sum_{j =1}^{4}
 \frac{1}{m_{\chi^+_k}}
\{\kappa_{\nu}|\tilde{D}^{\tau}_{1j}|^2 Re(D^{\nu \dagger}_{L_{11}}U_{k1}D^{\nu}_{R_{11}}V_{k2})\nonumber\\
+\kappa_N |\tilde{D}^\tau_{4j}|^2Re(D^{\nu \dagger}_{R_{12}}V^*_{k1}D^{\nu}_{L_{12}}U^*_{k2})\}
G_3(\frac{M_{\tilde{\tau}_j}}{m_{\chi^+_k}})\nonumber\\
+\frac{g^2m_em_{\nu_{\tau}}}{96\pi^2}\sum_{k =1}^{2} \sum_{j =1}^{4}
\frac{1}{m^2_{\chi^+_k}}
\{|\tilde{D}^\tau_{1j}|^2[|D^{\nu \dagger}_{L_{11}}U_{k1}|^2+\kappa^2_{\nu}|D^{\nu \dagger}_{R_{11}}V^*_{k2}|^2
]+\kappa^2_{\tau} |\tilde{D}^\tau_{3j}|^2|D^{\nu \dagger}_{L_{11}}U_{k2}|^2\nonumber\\
+|\tilde{D}^\tau_{4j}|^2[|D^{\nu \dagger}_{R_{12}}V^*_{k1}|^2+\kappa^2_{N}|D^{\nu \dagger}_{L_{12}}U_{k2}|^2
]+\kappa^2_{E_{\tau}}|\tilde{D}^\tau_{2j}|^2|D^{\nu \dagger}_{R_{12}}V^*_{k2}|^2\}
G_4(\frac{M_{\tilde{\tau}_j}}{m_{\chi^+_k}})
\eeqn
where 
\beqn
G_3(r)=\frac{-2}{r^2-1}+\frac{2r^2}{(r^2-1)^2}\ln(r^2),\nonumber\\
G_4(r)=\frac{3(1+r^2)}{(1-r^2)^2}+\frac{6r^2}{(1-r^2)^3}\ln(r^2).
\eeqn
 The numerical sizes of the neutrino moments $\mu_{\nu}^{(1)}$ and $\mu_{\nu}^{(2)}$ will be discussed
 in Sec.(5).

\section{$\tau$ anomalous magnetic moment}
 An evaluation of the  anomalous magnetic moment  in the standard model gives 
$a_{\tau}^{SM}=117721 (5) \times 10^{-8}$,
where $a_{\tau}=\frac{g_{\tau}-2}{2}$.
The experimental limits on this parameter are\cite{eidelman}
$-0.052<a_{\tau}^{exp}<0.013$
and so the sensitivity is more than
one order of magnitude   below where one can see the effects of the $\tau$ anomalous
magnetic moment.
 Here, we calculate the corrections to the $\tau$ anomalous magnetic moment including new physics
effects from the supersymmetrized mirror sector which mixes with the $\tau$ lepton sector. 
Specifically we compute 4 different types of loops corrections to $a_{\tau}$. These include the following 
exchanges in the loops:  (1) W boson and neutral mirror leptons; (2)
Z boson and charged mirror leptons ; (3) chargino and scalar neutrinos- mirror scalar neutrinos,
and (4)
neutralino, charged scalar leptons- mirror scalar leptons.
Using Eq.(\ref{LR}), one can write the contribution from the W boson loop so that 
\beqn
\Delta^{(1)} a_{\tau}=\frac{g^2}{8}\frac{m_{\tau}}{16\pi^2 M_W}
\sum_{\alpha,\gamma,\delta=1,2}
[|(D^{\nu\dagger}_{L})_{\alpha\gamma}g^L_{\gamma \delta}(D^{\tau}_{L})_{\delta 1}\nonumber\\
+(D^{\nu\dagger}_{R})_{\alpha\gamma}g^R_{\gamma \delta}(D^{\tau}_{R})_{\delta 1}|^2
-|(D^{\nu\dagger}_{L})_{\alpha\gamma}g^L_{\gamma \delta}(D^{\tau}_{L})_{\delta 1}
-(D^{\nu\dagger}_{R})_{\alpha\gamma}g^R_{\gamma \delta}(D^{\tau}_{R})_{\delta 1}|^2
]h_2(\frac{m_{\nu_{\alpha}}}{M_W}),
\eeqn 
where
\beq
h_2(r)=\frac{6r^5}{(r^2-1)^3}\ln r^2 +\frac{r^5-11r^3+4r}{(r^2-1)^2}.
\eeq
Using Eq.(\ref{zinteractions}), one can write the contribution from the Z boson loop
\beqn
\Delta^{(2)} a_{\tau}=\frac{g^2}{4\cos^2\theta_W}\frac{m_{\tau}}{16\pi^2 M_Z}
\sum_{j=1,2}
|x[-(D^{\tau \dagger}_L)_{j1}(D^{\tau}_L)_{11}\nonumber\\
+(D^{\tau \dagger}_R)_{j1}(D^{\tau}_R)_{11}
-(D^{\tau \dagger}_L)_{j2}(D^{\tau}_L)_{21}
+(D^{\tau \dagger}_R)_{j2}(D^{\tau}_R)_{21}]\nonumber\\
+\frac{1}{2}[(D^{\tau \dagger}_L)_{j1}(D^{\tau}_L)_{11}
-(D^{\tau \dagger}_R)_{j2}(D^{\tau}_R)_{21}]
|^2 h_1(\frac{m_{\tau_j}}{M_Z}),
\eeqn
where 
 $x$ is as  defined by Eq.(\ref{nc}) and 
\beq
h_1(r)=-\frac{6r^3}{(r^2-1)^3}\ln r^2 +\frac{r^5+r^3+4r}{(r^2-1)^2}.
\eeq

Next
using Eq.(\ref{charginoa}), one can write the contribution from the chargino, scalar
neutrino and scalar mirror neutrino as
\beqn
\Delta^{(3)} a_{\tau}=\frac{g^2m_{\tau}}{16\pi^2}\sum_{i =1}^{2} \sum_{j =1}^{4}
 \frac{1}{m_{\chi^+_i}}
\{\kappa_{\tau}|\tilde{D}^{\nu}_{1j}|^2 Re(D^{\tau \dagger}_{L_{11}}V_{i1}D^{\tau}_{R_{11}}U_{i2})\nonumber\\
+\kappa_{E_{\tau}} |\tilde{D}^\nu_{4j}|^2Re(D^{\tau \dagger}_{R_{12}}U^*_{i1}D^{\tau}_{L_{12}}V^*_{i2})\}
F_3(\frac{M^2_{\tilde{\nu}_j}}{m^2_{\chi^+_i}})\nonumber\\
+\frac{g^2m^2_{\tau}}{96\pi^2}\sum_{i =1}^{2} \sum_{j =1}^{4}
\frac{1}{m^2_{\chi^+_i}}
\{|\tilde{D}^\nu_{1j}|^2[|D^{\tau \dagger}_{L_{11}}V_{i1}|^2+\kappa^2_{\tau}|D^{\tau\dagger}_{R_{11}}U^*_{i2}|^2
]+\kappa^2_{\nu} |\tilde{D}^\nu_{3j}|^2|D^{\tau \dagger}_{L_{11}}V_{i2}|^2\nonumber\\
+|\tilde{D}^\nu_{4j}|^2[|D^{\tau \dagger}_{R_{12}}U^*_{i1}|^2+\kappa^2_{E_{\tau}}|D^{\tau \dagger}_{L_{12}}V_{i2}|^2
]+\kappa^2_{N}|\tilde{D}^\nu_{2j}|^2|D^{\tau \dagger}_{R_{12}}U^*_{i2}|^2\}
F_4(\frac{M^2_{\tilde{\nu}_j}}{m^2_{\chi^+_i}})
\eeqn
where
\beq
F_3(x)=\frac{1}{(x-1)^3}(3x^2-4x+1-2x^2 \ln{x}),
\eeq
and
\beq
F_4(x)=\frac{1}{(x-1)^4}(2x^3+3x^2-6x+1-6x^2 \ln{x}).
\eeq
Further,
using Eq.(\ref{neutralinoo}), one can write the contribution from the neutralino, scalar
lepton and scalar mirror lepton as
\beqn
\Delta^{(4)} a_{\tau}=-\frac{m_{\tau}}{32\pi^2}\sum_{k =1}^{4} \sum_{j =1}^{4}
 \frac{1}{m_{\chi^0_j}} F_1(\frac{M^2_{\tilde{\tau}_k}}{m^2_{\chi^0_j}})\nonumber\\
|\tilde{D}^{\tau}_{1k}|^2\{|(D^{\tau \dagger}_+)_{11}a_j+(D^{\tau \dagger}_-)_{11}b_j|^2
-|(D^{\tau \dagger}_-)_{11}a_j+(D^{\tau \dagger}_+)_{11}b_j|^2\}\nonumber\\
+|\tilde{D}^{\tau}_{4k}|^2\{|(D^{\tau \dagger}_+)_{12}a'_j-(D^{\tau \dagger}_-)_{12}b'_j|^2
-|(D^{\tau \dagger}_-)_{12}a'_j-(D^{\tau \dagger}_+)_{11}b'_j|^2\}\nonumber\\
+|\tilde{D}^{\tau}_{3k}|^2\{|(D^{\tau \dagger}_+)_{11}c_j+(D^{\tau \dagger}_-)_{11}d_j|^2
-|(D^{\tau \dagger}_-)_{11}c_j+(D^{\tau \dagger}_+)_{11}d_j|^2\}\nonumber\\
+|\tilde{D}^{\tau}_{2k}|^2\{|(D^{\tau \dagger}_+)_{12}c'_j-(D^{\tau \dagger}_-)_{12}d'_j|^2
-|(D^{\tau \dagger}_-)_{12}c'_j-(D^{\tau \dagger}_+)_{12}d'_j|^2\}\nonumber\\
+\frac{m^2_{\tau}}{96\pi^2}\sum_{k =1}^{4} \sum_{j =1}^{4}
 \frac{1}{m^2_{\chi^0_j}}F_2(\frac{M^2_{\tilde{\tau}_k}}{m^2_{\chi^0_j}})\nonumber\\
|\tilde{D}^{\tau}_{1k}|^2\{|(D^{\tau \dagger}_+)_{11}a_j+(D^{\tau \dagger}_-)_{11}b_j|^2
+|(D^{\tau \dagger}_-)_{11}a_j+(D^{\tau \dagger}_+)_{11}b_j|^2\}\nonumber\\
+|\tilde{D}^{\tau}_{4k}|^2\{|(D^{\tau \dagger}_+)_{12}a'_j-(D^{\tau \dagger}_-)_{12}b'_j|^2
+|(D^{\tau \dagger}_-)_{12}a'_j-(D^{\tau \dagger}_+)_{11}b'_j|^2\}\nonumber\\
+|\tilde{D}^{\tau}_{3k}|^2\{|(D^{\tau \dagger}_+)_{11}c_j+(D^{\tau \dagger}_-)_{11}d_j|^2
+|(D^{\tau \dagger}_-)_{11}c_j+(D^{\tau \dagger}_+)_{11}d_j|^2\}\nonumber\\
+|\tilde{D}^{\tau}_{2k}|^2\{|(D^{\tau \dagger}_+)_{12}c'_j-(D^{\tau \dagger}_-)_{12}d'_j|^2
+|(D^{\tau \dagger}_-)_{12}c'_j-(D^{\tau \dagger}_+)_{12}d'_j|^2\},
\eeqn
where
\beq
F_1(x)=\frac{1}{(x-1)^3}(1-x^2+2x \ln{x}),
\eeq
and 
\beq
F_2(x)=\frac{1}{(x-1)^4}(-x^3+6x^2-3x-2-6x \ln{x}).
\eeq
The numerical sizes of 
$\Delta^{(1)} a_{\tau}-  \Delta^{(4)} a_{\tau}$ are  discussed in in the next section.

\section{Constraints and Size Estimates}

There are severe phenomenological constraints on extra matter beyond the
Standard Model. These constraints can be listed as follows:
(1) constraints from the data on the Z width; (2) constraints from direct 
searches; (3) unitarity constraints on the  
enlarged $4\times 4$ CKM matrix; (4) constraints from the oblique 
electroweak effects; and (5) constraints  on Yukawas arising from keeping the 
theory perturbative, i.e., avoid developing a Landau pole.
Many of these constraints have been investigated  in the context of
a sequential fourth generation
\cite{Kribs:2007nz,Hung:2007ak,Holdom,Novikov:2002tk,Barger:1989dk,Frampton:1999xi,Carena:1995ep} with the analysis of \cite{Kribs:2007nz} being the
most recent and the most detailed. We summarize the main results of 
these analyses below. 
First of all the constraint (1) can be 
easily avoided by making the masses of the new particles greater than
half the Z boson width, while  (2) can be satisfied by putting lower bounds on new matter
from all collider data.
For example, the 
LEP II data puts bounds on charged
leptons of about a 100 GeV, while  the Tevatron puts bounds
on the fourth generation quark masses  so that
\cite{Kribs:2007nz}
that $m_{u_4} > 258$ GeV (95\% CL) and $m_{d_4} > 268$ GeV (at 95\% CL). 
(3) Regarding the CKM  unitarity constraints  the enlarged CKM matrix allows a small window
for mixings with the fourth generation so that \cite{Kribs:2007nz}
$|V_{14}|\leq .04$,  $|V_{41}|\leq .08$,  $|V_{24}|\leq .17$
and there are similar constraints  on the other mixings  
which allow for non-negligible elements for mixings with the 4th generation.\\

 Perhaps the most stringent of the constraint is (4) which comes from the 
oblique parameters $(S,T,U)$\cite{Peskin:1991sw,Altarelli:1991fk}
  and specfically from the oblique parameter $S$ (For a recent review of the S,T,U fits to
  the electroweak data see Ref.\cite{yao,lepweg}). 
  Here a complete fourth generation
  with degenerate masses
  gives a contribution of about 0.2.
However, this correction can be reduced when 
one considers splittings of the up and the down fermions in the same multiplet. 
Using such splittings  analyses  including the fourth generation allow for
consistent  $(S,T,U)$ fits to the data (see, e.g., \cite{Kribs:2007nz,Holdom}).  
(5) Finally it has been shown that  the Yukawa couplings  can remain perturbative up to the grand unification scale for a range of fourth generation masses and Higgs boson parameters.
Thus problems
such as generation of Landau pole singularities for a large 4th generation up quark mass can be avoided with  appropriate parameter choices.
Essentially all of the considerations valid for the sequential fourth generation are also valid for the a mirror generation. 
 Thus for, example, consider a fourth generation with up and down fermions $(\psi_1, \psi_2)$ with hypercharge $Y$
and masses $(M_1,M_2)$. 
The transformation that takes us 
 from fermions to mirror fermions is 
  \beqn
     {\rm fermions} ~(\psi_1, \psi_2)  \leftrightarrow  {\rm mirror ~fermions} ~(\psi^c_2, \psi^c_1),\nonumber\\
         Y   \leftrightarrow -Y,    M_1   \leftrightarrow  M_2. ~~~~~~~~~~~~~~~~~~
         \label{transformation}
          \eeqn
          Using the above one finds that 
      $\Delta S$ contribution from the  mirror generation is the same as for
 the 4th sequential  generation\cite{He:2001tp}.
  Without going into further details, we assume that fits to the electroweak data similar to those for
 the sequential fourth generation  can be carried out
 for the case of the  mirror generation. \\

Beyond the constraints on a new generation discussed above a mirror generation encounters two
more issues. The first concerns avoidance of the survival hypothesis\cite{Georgi:1979md}, i.e.,  a mirror generation and an
ordinary generation can combine to get super heavy masses of GUT size or string scale size.
However, it is well known that some of the mirror generations do escape gaining super heavy masses 
and remain light up to the electroweak scale\cite{Senjanovic:1984rw,Bagger:1984rk}.
 We assume in this analysis that this indeed is the case
for one mirror generation. The second issue concerns the mixing of the mirror generation with the 
 ordinary generations. In this work we assume that the mixing primarily occurs with the third
generation. In this circumstance the third generation will develop a small $V+A$ structure in addition
to the expected $V-A$ structure. 
Indeed 
such a $V+A$ component for some of the 
 third generation particles has been looked at for some time\cite{Jezabek:1994zv,Nelson:1997xd}.
 We here point out that the current data regarding the third generation leaves open the
 possibility of new physics.  
  For instance, the analysis of \cite{Choudhury:2001hs} finds a better fit to
 the precision electroweak data, and specifically a better fit to the forward-backward asymmetry
 $A_{FB}^b$ of the b -quark, with additional bottom like quarks. 
  Similarly, a model-independent  measurement of the W boson helicity in the top 
 quark decay $t\to Wb$ at D\O\ \cite{Abazov:2007ve},
  gives  for the longitudinal fraction $f_0$ and for the 
 right handed fraction $(f_+)$ the result  $f_0=.425\pm .166(\rm{stat}) \pm .102(\rm{syst})$  
  and $f_+=.119\pm .090(\rm{stat})\pm .053(\rm{syst})$ while $f_-$ is determined via
  the constraint $f_0+f_++f_-=1$. While the model independent analysis above is consistent with the 
  Standard Model prediction with $V-A$ structure of $f_0=.697, f_+=3.6\times 10^{-4}$, the
  analysis shows that a different Lorentz structure such as $V+A$  is not ruled out at the
  level of a few percent. A similar situation occurs in the analysis of $\tau$ lepton decays
  where new physics at the level of a few percent is not necessarily ruled out
 \cite{Dova:1998uj,Singh:1987hn}.\\

The mixing parameters and the
masses of the mirror fermion sector are determined by
the input parameters $\theta$, $\phi$, $m_N$ and $m_{E_{\tau}}$, where we
assume that $\theta_L=\theta_R=\theta$ and $\phi_L=\phi_R=\phi$ for the purpose
of numerical investigation.
However, these parameters are not independent but  constrained by the symmetry
breaking relation (\ref{condition}) which we use to  determine  $\phi$
in terms of the other parameters.
The scalar sector is determined by the 
mixing angles $\tilde{\theta}_{1,2}$ and $\tilde{\phi}_{1,2}$ 
and the simplifying assumption that the 
scalar (mass)$^2$  $4\times 4$ matrix factorizes into two $2\times 2$ block diagonal matrices.
If we further assume that $M^2_{i j}=M^2_{i+2 j+2}$ we  have
the conditions $\tilde{\theta}_1=\tilde{\theta}_2$ and
 $\tilde{\phi}_1=\tilde{\phi}_2$. The remaining parameters are
$M^2_{11}$ and $M^2_{22}$ for both the scalar $\tau$ and scalar neutrino 
(mass)$^2$ matrices. The scalar spectrum is then  calculated from the formulas given
in Appendix A.

\begin{center} \begin{tabular}{|c|c|c|c|c|c|c|c|c|}
\multicolumn{8}{c}{Table~1:  } \\
\hline
$\theta$ & $\tilde{\phi}$  &$\tilde{\theta}$ & $\Delta^{(1)} a_{\tau}$
 & 
$\Delta^{(2)} a_{\tau}$
  & $\Delta^{(3)} a_{\tau}$  &   $\Delta^{(4)} a_{\tau}$& 
$\mu^{(1)}_{\nu}$/$\mu_B$&  $\mu^{(2)}_{\nu}$/$\mu_B$  \\
 & & & $\times 10^{6}$ &$\times 10^{7}$ &$\times 10^{7}$ &$\times
 10^{8}$ 
&$\times 10^{10}$&$\times 10^{10}$  \\
\hline
$0.2$     &  $0.3$  &  $0.4$     &   $5.0 $    &    $18.$   &   $2.4$
    
&   $-8.1 $  &  $-24.$ & $15. $\\
 \hline
$0.15$     &  $0.35$  &  $0.45$     &   $2.8$    &    $10. $   &  
 $1.4 $    
&   $-4.8 $  &  $-14. $ & $8.7 $\\
\hline
 $0.10$     &  $0.2$  &  $0.3$     &   $1.3 $    &    $4.7 $   &  
 $.59$    
&   $-1.92 $  &  $-6.2 $ & $3.8 $\\
 \hline
$0.09$     &  $0.0$  &  $0.2$     &   $1.06$    &    $3.8.$   &   $.47
 $    
&   $-1.52 $  &  $-4.90 $ & $3.1 $\\
 \hline
 $0.08$     &  $0.2$  &  $0.1$     &   $.84$    &    $3.0$   &   $
 .38$    
&   $-1.19 $  &  $-3.95$ & $2.4 $\\
 \hline
$0.07$     &  $0.1$  &  $0.0$     &   $.65$    &    $2.30 $   &   $.29
 $    
&   $-.91 $  &  $-3.04$ & $1.8 $\\
 \hline
$0.06$     &  $0.0$  &  $0.2 $     &   $.48$    &    $1.70 $   &  
 $.21 $    
&   $-.67 $  &  $-2.23 $ & $1.4 $\\
 \hline
$0.05$     &  $0.2$  &  $0.1$     &   $.33$    &    $1.18 $   &   $.15
 $    
&   $-.64$  &  $-1.55 $ & $.94 $\\
 \hline
 $0.04$     &  $0.1$  &  $0.0$     &   $.21$    &    $.76$   &   $.09
 $    
&   $-.30$  &  $-.99 $ & $.60 $\\
 \hline
 $0.03$     &  $0.0$  &  $0.2$     &   $.12 $    &    $.43 $   &  
 $.05 $    
&   $-.17$  &  $-.56 $ & $.34$\\
 \hline
  $0.02$     &  $0.2$  &  $0.1$     &   $.05 $    &    $.19$   &  
 $.03$    
&   $-.07 $  &  $-.25$ & $.15 $\\
 \hline
   $0.01$     &  $0.1$  &  $0.0$     &   $.013 $    &    $.048 $   &  
 $.006$    &   $-.02  $  &  $-.062 $ & $.037 $\\
 \hline
\end{tabular}\\~\\
\noindent
\end{center}
Table caption: Contributions to the magnetic moments of  $\nu_{\tau}$
 and of $\tau$  
including  corrections from the mirror particles and mirror sparticles
 for a variety of mixing angles
between the third generation and the mirror generation consistent with
 the symmetry
breaking constraint of Eq.(\ref{condition}).
The other input parameters are $\tan\beta=20$, $m_0=400$,
 $m_{1/2}=150$,
$A_0=400$,  $m_E=200$,
 $m_N=220$,
 $M_{\tilde{\tau}_{11}}=400$,
$M_{\tilde{\tau}_{22}}=500$, $m_{\tilde{\nu}_{11}}=420$ and
 $m_{\tilde{\nu}_{22}}=520$,
and  $\mu>0$.
All masses are in units of GeV and all angles are in radian.\\

The mixings between the third generation and the mirrors can affect among other things
the magnetic moments. This is specifically true for the magnetic moment of the $\tau$ 
neutrino which we discuss next. 
In this case there will be two contributions,
one from the non-susy sector (see Fig.(1)) and the other from the SUSY sector (see Fig.(2)). 
Similar contributions also arise for the anomalous magnetic moment of 
the $\tau$. An analysis of these moments is given in  Table 1. 
Here we exhibit numerical sizes of 
the different  contributions to the tau neutrino
magnetic moments, i.e., $\mu_{\nu}^{(1)}$ and $ \mu_{\nu}^{(2)}$ and to the 
anomalous magnetic moment of the $\tau$, i.e., $\Delta^{(1)} a_{\tau}- 
\Delta^{(4)} a_{\tau}$. 
The numerical results of the table show  that the contribution to the 
$\tau$ neutrino magnetic moment is as much as eight  orders of 
magnitude larger than what the model without mirror mixings will give.
These results may be compared with the prediction of the Standard Model
(extended with a right handed neutrino) which is $\mu_{\nu}=O(10^{-19}) (m_{\nu}/eV)\mu_B$.
The SM  value for the magnetic moment is too small and falls  beyond any reasonable
possibility of observation.
In contrast the result arising from mixing with the mirror sector is only 2-3 orders of
magnitude 
below the current limits and thus not  outside the realm of observability.
At the same time, we note that the contribution of the 
 mirror sector to the anomalous magnetic moment of the $\tau$ lepton
  gives only a small correction to the Standard Model prediction.

\section{LHC Signatures of  the mirror sector\label{sig}}
Before  discussing the LHC signatures of the mirror sector it is useful to list the new 
particles that arise in the model beyond those that appear in MSSM. 
In the fermionic sector the new particles are 
\beqn
B, T, E, N
\label{mirror-1}
\eeqn
where all fields including $N$ are Dirac. 
In the bosonic sector the new particles in the mass diagonal states  are 
\beqn
\tilde B_1, \tilde B_2, \tilde T_1, \tilde T_2, \tilde E_1, \tilde E_2, 
~\tilde \nu_{1}, ~\tilde \nu_{2},
~\tilde \nu_{3}.
\label{mirror-2}
\eeqn
We note the appearance of three sneutrino states in Eq.(\ref{mirror-2}). 
This is so because, we started out with
two extra chiral singlets, one in the MSSM sector and another in the mirror generation.
Along with the two chiral neutrino states that arise  from the doublets they produce 
four sneutrino states, one of which is in the MSSM sector and the other three 
are listed in Eq.(\ref{mirror-2}).\\

In the extended MSSM with mirrors, 
the mirror fermions  and their supersymmetric partners, the mirror
sfermions,
could produce interesting signatures at the LHC and at the ILC. 
Thus, for example, if the mirror generation mixes only with the 
third  generation one will have decays of the following type 
(if $M_{N}> M_E+ M_W$),
\beqn
N\to E^-W^+, ~
E^-\to \tau^-Z\to \tau^-e^+e^-, \tau^-\mu^+\mu^-, \tau^+\tau^+\tau^-
\eeqn
This signal is unique in the sense that there is always
at least one $\tau$. Specifically,  there is
no corresponding signal where one has all three leptons of the first 
generation, or of the second generation or a mixture there of.  
These signatures are uniquely  different from the leptonic signatures 
in MSSM, for example, from those 
arising from the decay of an off -shell $W^*$\cite{Nath:1987sw},  
where $\tilde W^*\to \tilde W +\chi_2^0\to l_1  l_2\bar l_2$, i.e.,with a $W^*$
decaying into 
a chargino and the second lightest neutralino.
  Here  all leptonic
generations appear in all final states.  Another interesting signature
is the Drell-Yan process 
\beqn 
p p \to Z^* \to E^+E^-  \to 2\tau 4l, 4\tau 2l, 6\tau, 
\eeqn
where $l_1,l_2= e, \mu$. Additionally,  of course,  there can be events 
with  taus, leptons and jets.  
In each case one has two opposite sign taus. Similarly one can have
 $pp\to Z^*\to  N\bar N$
 production.
One can also have the production of mirrors via $W^*$ exchange, i.e., via  the process 
\beqn
pp\to W^*\to EN
\to  [\tau l_i\bar l_i,  3\tau, (\tau +2 jets)] + E_T^{\rm miss}
\eeqn
Again the leptonic events always have a $\tau$ with no events of the type $l_1l_2\bar l_2$. 
 Similarly decay chains exist with other mass hierachies, e.g.,
when $N$ 
is lighter than $E$.  
Additionally for the supersymmetric sector of mirMSSM one has production and decays
of $\tilde E_{1,2}$ and $\tilde \nu_{i}$ (i=1,2,3).  For example, for the case, when
$\tilde \nu_{i}$ are heavier than $\tilde{E_k}$ one has decays 
\beqn
\tilde \nu_{i} \to  \tilde E_k^- W^+, E^-\tilde \chi^+ 
\eeqn
with subsequent decays of $E^-,\tilde E_k^-$ etc. 
Thus one has processes of the type 
\beqn
pp \to 
 \tilde \nu_{i} {\tilde \nu^*_{i}} 
 \to 
\tilde E^+_k \tilde E^-_kW^+W^-, 
\tilde E^+_kE^-W^{\mp} \tilde \chi^{\pm}
\eeqn
Combined with the decays of the $\tilde E^+\tilde E^-$
one can get 
 signatures with $\tau s +{\rm  leptons} + {\rm jets} + E_T^{\rm miss}$
 with as many 8 leptons,  where all the leptons could be $\tau$s. 
 Another important signature is the radiative decay\cite{De Rujula:1980qd}
  of  $N$ where 
 \beqn
 N\to  \nu_{\tau} \gamma.
 \eeqn
 This decay occurs via the transition electric and magnetic moments. The lifetime  for the
 decay is very short and once $N$
 is produced it will decay inside the detector.  The signal will consist of a very energetic
 photon with energy in the 100 GeV range.  
 Thus if kinematically allowed 
 $h^0, A^0$ will have decays of the following typess
 \beqn
(h^0, H^0, A^0)\to  N\bar N \to 2\gamma+ E_T^{\rm miss}. 
\eeqn
 Once  a new generation is seen, a study of their production and decay
   can reveal if they are a sequential generation or a mirror generation. Let us consider
   the sequential fourth generation first with the superpotential
\beqn
W_{4th-seq}= \epsilon_{ij}[y_{4e}\hat H^i_1 \hat \psi^j_{4L} \hat e^c_{4L}
+y_{4d} \hat H^i_1 \hat q^j_{4L} \hat d^c_{4L} 
+y_{4u} \hat H^j_2 \hat q^i_{4L} \hat u^c_{4L}
++y_{4\nu} \hat H^j_2 \hat \psi^i_{4L} \hat \nu^c_{4L}]
\eeqn
      which  relate the Yukawas with the 
    fermion masses  for the 4th generation  so that 
    \beqn
    y_{4u}= \frac{g m_{4u}}{\sqrt 2 M_W\sin\beta},
y_{4\nu}= \frac{g m_{4\nu}}{\sqrt 2 M_W\sin\beta},\nonumber\\
        y_{4e}= \frac{g m_{4e}}{\sqrt 2 M_W\cos\beta}, 
     ~y_{4d}= \frac{g m_{4d}}{\sqrt 2 M_W\cos\beta}.
     \eeqn
    For the mirror generation we have
\beqn
W_{4th-m}=\epsilon_{ij}[f_2 \hat H_1^i \hat \chi^{cj} \hat N_L +f_2' \hat H_2^j \hat \chi^{ci} \hat E_{\tau L}
+Y_B \hat H_2^j \hat Q^{ci} \hat B_L +Y_T \hat H_1^i \hat Q^{cj} \hat T_L]
\eeqn
         and  the relation among the Yukawas and the mirror fermions masses are 
   \beqn
   f_2= \frac{g M_{N}}{\sqrt 2 M_N\cos\beta},    ~~Y_{T}= \frac{g M_{T}}{\sqrt 2 M_W\cos\beta},\nonumber\\
        f_2'= \frac{ M_{E}}{\sqrt 2 M_W\sin\beta},
~Y_{B}= \frac{g M_{B}}{\sqrt 2 M_W\sin\beta}.
\eeqn
   The neutral Higgs mass eigen states $h^0$, $H^0$ and $A^0$ are related to the electroweak eigen states $H_1^1$ and $H_2^2$ by
\beqn
H_1^1=\frac{1}{\sqrt{2}}[v_1+H^0 \cos\alpha -h^0 \sin\alpha +i A^0 \sin\beta]\nonumber\\
H_2^2=\frac{1}{\sqrt{2}}[v_2+H^0 \sin\alpha +h^0 \cos\alpha +i A^0 \cos\beta]
\eeqn

The neutral Higgs couplings of $h^0, H^0$ and of the CP odd Higgs boson $A^0$  with  the sequential 4th generation 
in the Lagrangian 
takes  the form
\beqn
-{\cal{L}}=\frac{g}{2M_W}
(\frac{ m_{4e}\cos\alpha}{ \cos\beta} \bar e_4 e_4 + 
\frac{ m_{4d}\cos\alpha}{  \cos\beta} \bar d_4 d_4+ 
\frac{ m_{4u}\sin\alpha}{  \sin\beta}  \bar u_4u_4+
\frac{ m_{4\nu}\sin\alpha}{  \sin\beta}  \bar \nu_4 \nu_4)H^0\nonumber\\
+ \frac{g}{2M_W}
(-\frac{ m_{4e}\sin\alpha}{  \cos\beta} \bar e_{4}e_4 
-\frac{ m_{4d}\sin\alpha}{ \cos\beta} \bar d_{4}d_4 + 
\frac{ m_{4u}\cos\alpha}{  \sin\beta} \bar  u_{4}u_4
+\frac{ m_{4\nu}\cos\alpha}{  \sin\beta} \bar  \nu_{4}\nu_4)h^0\nonumber\\
-\frac{ig}{2M_W}(m_{4e}\bar e_4\gamma_5e_4 \tan\beta + m_{4d}
\bar d_4 \gamma_5 d_4 \tan\beta
+ m_{4u} \bar u_4 \gamma_5 u_4 \cot\beta
+m_{4\nu} \bar \nu_4 \gamma_5 \nu_4 \cot\beta) A^0,
\label{h01}
\eeqn
while for the mirror generation it takes  the form
\beqn
-{\cal{L}}=\frac{g}{2M_W}
(\frac{ M_{E}\sin\alpha}{  \sin\beta} \bar E E + 
\frac{ M_{B}\sin\alpha}{ \sin\beta} \bar B B  +
\frac{ M_{T}\cos\alpha}{ \cos\beta} \bar T T+
\frac{ M_{N}\cos\alpha}{ \cos\beta} \bar N N
)H^0\nonumber\\
+\frac{g}{2M_W}
(\frac{ M_{E}\cos\alpha}{ \sin\beta} \bar E E +
\frac{ M_{B}\cos\alpha}{ \sin\beta} \bar B B -
\frac{ M_{T}\sin\alpha}{ \cos\beta} \bar T  T-
 \frac{ M_{N}\sin\alpha}{ \cos\beta} \bar N N
)h^0\nonumber\\
-\frac{ig}{2M_W}(M_{E}\bar E\gamma_5E \cot\beta + M_{B}
\bar B \gamma_5 B \cot\beta
+ M_{T} \bar T \gamma_5 T \tan\beta
+ M_{N} \bar N \gamma_5 N \tan\beta
) A^0.
 \label{h02}
\eeqn
A comparison of Eq.(\ref{h01}) and of Eq.(\ref{h02}) shows a rearrangment of
$\alpha$ and $\beta$ dependence. Thus while the down quark and the lepton vertices for 
a sequential generation are enhanced for large $\tan\beta$, it is the up quark vertex for
a mirror generation  that is enhanced. The above leads to some interesting features that
distinguish a mirror generation from a sequential fourth generation.

One important consequence of the above is the following. Suppose the $H^0$ is heavy
enough to decay into a pair of fourth generation quarks or a pair of mirror quarks
($m_{H^0}>2m_q, q=u_4, d_4$).
Then let us define the  ratio of branching ratios $R_{d_4/u_4}^{H^0}$ as 
\beqn
R_{d_4/u_4}^{H^0}= BR(H^0\to d_4\bar d_4)/ BR(H^0\to u_4\bar u_4).
\eeqn
Using the vertices in Eq.(\ref{h01}) we find
\beqn
R_{d_4/u_4}^{H^0}= \frac{m_{d_4}^2}{m_{u_4}^2} (\cot\alpha\tan\beta)^2 P_{d_4/u_4}^{H^0},
\eeqn
where $P_{d_4/u_4}^{H^0}$ is a phase space factor defined by 
$P_{d_4/u_4}^{H^0}=(1-4m_{d_4}^2/m_H^2)^{3/2}(1-4m_{u_4}^2/m_H^2)^{-3/2}$ (see Appendix B). 
Similarly if the heavy Higgs can decay into the mirror quarks ($m_{H^0}>2m_Q, Q=B,T$)
one has 
\beqn
R_{B/T}^{H^0}= \frac{m_{B}^2}{m_{T}^2} (\tan\alpha\cot\beta)^2 P_{B/T}^{H^0},
\eeqn
where we have neglected  the loop effects. Thus with a 
knowledge of the parameters of the Higgs sector, i.e., $\alpha$ and $\beta$ one has
a way of differentiating a mirror generation from a sequential fourth generation.
Even a more dramatic differentiation arises from the branching ratios involving the decay
of the CP odd Higgs. Here one finds 
\beqn
R_{d_4/u_4}^{A^0} =\frac{m_{d_4}^2}{m_{u_4}^2} \tan^4\beta P_{d_4/u_4}^{A^0},
\eeqn
where 
$P_{d_4/u_4}^{A^0}=(1-4m_{d_4}^2/m_A^2)^{1/2}(1-4m_{u_4}^2/m_A^2)^{-1/2}$
while a similar ratio for the decay into the mirror quarks gives (see Appendix B)
\beqn
R_{B/T}^{A^0} = \frac{m_{B}^2}{m_{T}^2} \cot^4\beta P_{B/T}^{A^0},
\eeqn
where again we have neglected possible loop effects. 
The above implies that for $\tan\beta\geq 2$, $A^0$ will dominantly decay into $d_4\bar d_4$
for the sequential fourth generation case, while it will decay dominantly into $T\bar T$ 
for  a mirror generation. 
 Another important way to discriminate between a sequential generation and a mirror generation is
to look  at the forward backward asymmetry. 
Thus for the process $f\bar f \to f' \bar f'$  one may define,  the forward-backward asymmetry 
$A_{FB}$ = $( \int_0^1 dz (d\sigma/dz)$ - $\int_{-1}^0 dz (d\sigma/dz))$ $/$ 
$(\int_{-1}^1 dz (d\sigma/dz))$.  This asymmetry is sensitive to the $V+A$
vs $V-A$ structure of the $f'$ fermion interaction and a measurement of it can help
discriminate between a sequential generation and  a mirror generation. 
In  the above we have given a broad outline of the ways in which one might distinguish a mirror generation from a sequential fourth  generation.
There are many other possible chains for decay of the mirrors and mirror sparticles depending
on their  mass patterns. Further, 
  more detailed analyses of signatures for the model with mirrors based on detector simulations would
  be useful   along
the line of the analysis of  signatures for sugra models\cite{msugra} 
and for string models (For, a sample of recent works see\cite{Feldman:2007zn,kks,arnowitt,mmt,bps}). 
Finally  we comment on the flavor changing neutral current (FCNC) issues. It is well known
that mixing with mirrors frustrates the GIM mechanism which suppresses FCNC. 
For the  current model this does not pose a problem because the mirrors  do not mix with 
the first two generations. On the other hand one does have  couplings of the $Z$ boson which
are off diagonal, $Z\bar \tau E$, $Z \bar b B$, $Z\bar t T$ etc which would allow
production via a Drell -Yan process of  $pp\to Z^*\to \tau^+E^-, t \bar T, b\bar B$ etc,
which are not allowed for a sequential generation. Of course the processes are suppressed
by mixing angles.

\section{Conclusion\label{conclusion}}
In this work we consider  an extension of MSSM with an extra mirror
generation which remains light down to the electroweak scale. 
Recent analyses indicate that  an extra sequential generation is not inconsistent with
the precision electroweak data, and similar considerations apply to a mirror
generation. In the  model we consider, we allow for mixings of the mirror generation 
with the third generation, and investigate some of the phenomenological 
implications of the model. One important effect arises on the magnetic moment of
the $\tau$ neutrino, where one finds that
it is enhanced
by up to eight  to nine orders of magnitude over  what is predicted in the Standard Model.
We also discussed the possible signatures of the mirror generation at the LHC,
and find that several characteristic signatures exist which would distinguish it
from a sequential generation. One such crucial  test is the measurement of the 
forward -backward asymmetry which can discriminate between the $V-A$ vs
$V+A$ interactions. It is further shown that the couplings of the mirror generation
have different $\tan\beta$ dependences than those of an ordinary generation
or of  a sequential 4th generation.\\

 If  a  mirror generation exists,  it  has important 
implications for string model building. (For some recent work in D brane and  string model
building see \cite{Blumenhagen:2001te,Cvetic:2001nr,Kobayashi:2004ya,Bouchard:2005ag,Braun:2005nv,Lebedev:2007hv}).  
Typically in string model building one puts in the constraints that the difference between
the number of generations $n_f$ and the 
mirror generations $n_{mf}$ (with $n_f >n_{mf})$ equal three. 
This assumes that the $n_{mf}$  number of generations
 and mirror generations follow the survival hypothesis \cite{Georgi:1979md}
 and become superheavy. However, in unified models there
are many instances where mirror generations may remain massless up to the electroweak scale.
This opens a new direction for model building. 
Suppose, then, that one imposes only the constraint 
  $n_f-n_{mf}=2$ along 
  with the condition that one  mirror  generation remains massless
down to the electroweak scale. In this case we will have three ordinary generations and one
mirror generation all light at the electroweak scale, i.e., the extended MSSM
model with mirrors.\\

If  the scenario outlined above holds, the string model 
building may need a revision in that  the constraint of three massless generations will be relaxed. 
Specifically, for example,  in Kac-Moody level 2 
heterotic string constructions one has  problems getting 3 massless
generations (see,e.g., \cite{Kakushadze:1996jm}).
On the other hand, if 3 ordinary generations and one mirror generations are
massless, the rules of construction for string models change and one
may need to take a fresh look at model building in string theory. 
 Of course, the light  mirror particles even if they exist 
need not necessarily fall into a full generation. Thus while a full generation is the simplest
possibility for the cancellation of anomalies, it may happen that such cancellations may 
involve some exotic mirrors. This would make model building even more challenging. 
Many open question remain for further study 
the most important of which is a detailed 
dynamical model for the mixings of ordinary and mirror particles below the
grand unification scale.
In the analysis given in this work we assumed a phenomenological approach where
we introduce mixings between the two sectors. However, a concrete mechanism 
is desirable to achieve a more complete understanding of the mixings of the  ordinary
matter  and mirror matter.

\noindent
{\large\bf Acknowledgments}\\ 
Interesting conversations with Emanuela Barberis, 
Patrick Huber, Stuart Raby and Akin Wingerter are acknowledged.
This research is  supported in part by NSF grant PHY-0757959. 

\section{Appendix A: Further details of   mixings and interactions}
In this section we give  more explicit forms for the interactions including
mixing with mirrors.
We first discuss the non-supersymmetric sector where the contributions arise from the 
W and Z  exchanges.  
By parametrizing the mixing between $\tau$ and $E_{\tau}$ by the angle $\theta$, and between $\nu$ and $N$
by the angle $\phi$, in the simple case where $\theta_L=\theta_R=\theta$ and
$\phi_L=\phi_R=\phi$, we can write ${\cal{L}}_{CC}+{\cal{L}}_{NC}$ as 
\beqn
{\cal{L}}_{CC}+{\cal{L}}_{NC}= -\frac{g}{2\sqrt 2} W_{\mu}^{\dagger} 
\{\bar \nu_1\gamma^{\mu} \tau_1 \cos(\theta-\phi) 
+\bar \nu_1 \gamma^{\mu} \tau_2 \sin(\theta-\phi) \nonumber\\
-\bar \nu_1\gamma^{\mu} \gamma_5 \tau_1 \cos(\theta+\phi) 
-\bar \nu_1 \gamma^{\mu} \gamma_5 \tau_2\sin(\theta+\phi)\nonumber\\
-\bar \nu_2\gamma^{\mu} \tau_1 \sin(\theta-\phi) 
-\bar \nu_2 \gamma^{\mu}\gamma_5 \tau_1 \sin(\theta+\phi) \nonumber\\
+\bar \nu_2\gamma^{\mu}  \tau_2 \cos(\theta-\phi) 
+\bar \nu_2 \gamma^{\mu} \gamma_5 \tau_2\cos(\theta+\phi)\}+H.c\nonumber\\
-\frac{g}{4\cos\theta_W}Z_{\mu}
\{\bar \tau_1\gamma^{\mu} (4\cos^2\theta_W -1+\cos 2\theta \gamma_5)\tau_1 \nonumber\\
+\bar \tau_2\gamma^{\mu} (4\cos^2\theta_W -1-\cos 2\theta \gamma_5)\tau_2\nonumber\\
+\bar \tau_1\gamma^{\mu}\gamma_5 \sin 2\theta \tau_2+
\bar \tau_2\gamma^{\mu}\gamma_5 \sin 2\theta \tau_1\},
\label{a1}
\eeqn
where $\tau_1, \tau_2$ are the mass eigen states for the charged leptons, with 
$\tau_1$ identified  as the physical tau state, 
and $\nu_1, \nu_2$ are the mass eigen states for the neutrino
 with $\nu_1$ identified as the observed 
neutrino. We note  that Eq.(\ref{a1}) conicides with  Eq.(1) 
of \cite{mirrors} except for the typo in the middle sign of their third line.\\
 
 In the supersymmetric sector,
the mass terms of the scalar leptons and scalar mirror leptons arise from the F-term, the D-term and 
the soft supersymmetry breaking terms in the scalar potential. For example, the mixing terms
between $\tilde{\tau}_L$ and $\tilde{\tau}_R$ can arise from the $\mu$ term in the superpotenital and from
the trilinear coupling term of the soft breaking potential $V_{soft}$. This gives us 
the  terms $M^2_{13}=M^2_{31}=m_{\tau}(A_{\tau}-\mu \tan\beta)$. The corresponding mixing terms
between $\tilde{E}_{\tau L}$ and $\tilde{E}_{\tau R}$ are $M^2_{24}=M^2_{42}=m_{E_{\tau}}(A_{E_{\tau}}-\mu \cot\beta)$.
We assume here that the couplings are real otherwise,
we would have $M^2_{31}=m_{\tau}(A^*_{\tau}-\mu^* \tan\beta)$
and $M^2_{42}=m_{E_{\tau}}(A^*_{E_{\tau}}-\mu^* \cot\beta)$.
In the general parameter space of MSSM one can fix these mixings to be zero by a proper choice of the parameters
$\mu$, $A_{\tau}$ and $A_{E_{\tau}}$. The other
 elements of the scalar mass$^2$ matrix can also be easily
  worked out. As an example, the F-term 
produces a part of the
 mixing between $\tilde{\tau}_R$ and $\tilde{E}_{\tau R}$ as follows
\beq
V=F_i^* F_i,~ F_i=\frac{\partial W}{\partial A_i}.
\eeq 
Here $A_i$ is the scalar $\tilde{E}_{\tau L}$ and 
\beq
\frac{\partial W}{\partial \tilde{E}_{\tau L}}=f_2' H_2^2 \tilde{E}^*_{\tau R}+f_4 \tilde{\tau}^*_R-f_2'H_2^1 \tilde{N}^*_R,
\eeq 
which gives
\beq
V_F=(f_2' H_2^{2*} \tilde{E}_{\tau R}+f_4 \tilde{\tau}_R-f_2'H_2^{1*} \tilde{N}_R)(f_2' H_2^2 \tilde{E}^*_{\tau R}+f_4 \tilde{\tau}^*_R-f_2'H_2^1 \tilde{N}^*_R).
\eeq
 After breaking of the electroweak symmetry the $V_F$ part of the scalar potential given above 
 produces the following mass terms
\beq
-{\cal{L}}_{m}=f^{'2}_2 \frac{v^2_2}{2}\tilde{E}_{\tau R} \tilde{E}^{*}_{\tau R}
+f_4 f_2' \frac{v_2}{\sqrt{2}}\tilde{E}^{*}_{\tau R} \tilde{\tau}_R
+f_4 f_2' \frac{v_2}{\sqrt{2}}\tilde{E}_{\tau R} \tilde{\tau}^{*}_R
+f^2_4 \tilde{\tau}^{*}_R \tilde{\tau}_R
\eeq
 Here one finds that
 the mixing between $\tilde{\tau}_R$ and $\tilde{E}_{\tau R}$ occurs such that
the corresponding elements in the mass$^2$ matrix $M^2_{34}$ and $M^2_{43}$ are equal.
 
For illustrative purposes, we assume a simple mixing 
 scenario for mixings in the scalar  sector. Specifically we assume mixings among 
 scalars and mirror scalars of the same chirality. Thus for the charged leptons we assume
 mixings 
  between $\tilde \tau_L$ and $\tilde E_L$ and 
 similarly mixings between $\tilde \tau_R$ and $\tilde E_R$, but no 
 mixing between $\tilde \tau_L, \tilde \tau_R$
 and between $\tilde E_L$ and  $\tilde E_R$. These are obviously approximations to the
 more general analysis given in Sec.(2). 
 Under the above approximations  the diagnolizing matrices
$\tilde{D}^{\tau}$ and $\tilde{D}^{\nu}$ would have the following simple structures
\beqn
\tilde{D}^{\tau}={\left(
\begin{array}{cccc}
\cos\tilde{\theta}_1 & \sin\tilde{\theta}_1 &0&0 \cr
       -\sin\tilde{\theta}_1 & \cos\tilde{\theta}_1 &0&0\cr
0&0&\cos\tilde{\theta}_2 & \sin\tilde{\theta}_2\cr
0&0&-\sin\tilde{\theta}_2 & \cos\tilde{\theta}_2
 \end{array}\right)},
\eeqn
and
\beqn
\tilde{D}^{\nu}={\left(
\begin{array}{cccc}
\cos\tilde{\phi}_1 & \sin\tilde{\phi}_1 &0&0 \cr
             -\sin\tilde{\phi}_1 & \cos\tilde{\phi}_1 &0&0\cr
0&0&\cos\tilde{\phi}_2 & \sin\tilde{\phi}_2\cr
0&0&-\sin\tilde{\phi}_2 & \cos\tilde{\phi}_2
 \end{array}\right)}.t
\eeqn
In the charged leptonic sector, assuming the independent set of parameters to be 
$\tilde{\theta}_1$, $\tilde{\theta}_2$, $M^2_{11}$,
$M^2_{22}$, $M^2_{33}$ and $M^2_{44}$, one
can determine the elements $|M^2_{12}|$ and $|M^2_{34}|$ through the relations
\beqn
\tan 2\tilde{\theta}_1=\frac{2|M^2_{12}|}{M^2_{11}-M^2_{22}},\nonumber\\
\tan 2\tilde{\theta}_2=\frac{2|M^2_{34}|}{M^2_{33}-M^2_{44}}.
\eeqn
The eigen values for the masses are then given by 
\beqn
M^2_{{\tilde{\tau}}_1}=\frac{1}{2}(M^2_{11}+M^2_{22})+\frac{1}{2}\sqrt{(M^2_{11}-M^2_{22})^2+4|M^2_{12}|^2},\nonumber\\
M^2_{{\tilde{\tau}}_2}=\frac{1}{2}(M^2_{11}+\frac{1}{2}M^2_{22})-\frac{1}{2}\sqrt{(M^2_{11}-M^2_{22})^2+4|M^2_{12}|^2},\nonumber\\
M^2_{{\tilde{\tau}}_3}=\frac{1}{2}(M^2_{33}+M^2_{44})+\frac{1}{2}\sqrt{(M^2_{33}-M^2_{44})^2+4|M^2_{34}|^2},\nonumber\\
M^2_{{\tilde{\tau}}_4}=\frac{1}{2}(M^2_{33}+M^2_{44})-\frac{1}{2}\sqrt{(M^2_{11}-M^2_{44})^2+4|M^2_{34}|^2}.
\eeqn
Similar relations hold for the scalar neutrino sector.

\section{Appendix B: Decay of the heavy Higgs Bosons $H^0$ and $A^0$ into mirrors}
The heavy Higgs decays into mirrors would produce some very characteristic 
signatures if the masses of the heavy Higgs bosons $H^0$ and $A^0$ are large
enough to kinematically allow such decays. We give below the  decay widths for
the processes with charged mirrors
\beqn
H^0\to  E\bar E, B\bar B, T\bar T,\nonumber\\
A^0\to  E\bar E, B\bar B, T\bar T,
\eeqn 
using the interactions of Eq.(\ref{h02}). For the decay of $H^0$ into charged mirrors we have
\beqn
\Gamma(H^0\to E\bar E)= \frac{g^2m_{H^0}}{32\pi} (\frac{\sin\alpha}{\sin\beta})^2 (\frac{M_E}{M_W})^2
(1-\frac{4M_E^2}{M_H^{02}})^{3/2}, \nonumber\\
\Gamma(H^0\to B\bar B)= \frac{3g^2m_{H^0}}{32\pi} (\frac{\sin\alpha}{\sin\beta})^2 (\frac{M_B}{M_W})^2
(1-\frac{4M_B^2}{M_H^{02}})^{3/2}, \nonumber\\
\Gamma(H^0\to T\bar T)= \frac{3g^2m_{H^0}}{32\pi} (\frac{\cos\alpha}{\cos\beta})^2 (\frac{M_T}{M_W})^2
(1-\frac{4M_T^2}{M_H^{02}})^{3/2}.
\eeqn
These may be compared with the decays of $H^0$ into a 4-th sequential generation which are
\beqn
\Gamma(H^0\to e_4\bar e_4)= \frac{g^2m_{H^0}}{32\pi} (\frac{\cos\alpha}{\cos\beta})^2 (\frac{m_{e_4}}{M_W})^2
(1-\frac{4m_{e_4}^2}{M_H^{02}})^{3/2}, \nonumber\\
\Gamma(H^0\to d_4\bar d_4)= \frac{3g^2m_{H^0}}{32\pi} (\frac{\cos\alpha}{\cos\beta})^2 (\frac{m_{d_4}}{M_W})^2
(1-\frac{4m_{d_4}^2}{M_H^{02}})^{3/2}, \nonumber\\
\Gamma(H^0\to u_4\bar u_4)= \frac{3g^2m_{H^0}}{32\pi} (\frac{\sin\alpha}{\sin\beta})^2 (\frac{m_{u_4}}{M_W})^2
(1-\frac{4m_{u_4}^2}{M_H^{02}})^{3/2}.
\eeqn
For the decay of $A^0$ into charged mirrors we have
\beqn
\Gamma(A^0\to E\bar E)= \frac{g^2m_{A^0}}{32\pi} \cot^2\beta (\frac{M_E}{M_W})^2
(1-\frac{4M_E^2}{M_A^{02}})^{1/2}, \nonumber\\
\Gamma(A^0\to B\bar B)= \frac{3g^2m_{A^0}}{32\pi}   \cot^2\beta  (\frac{M_B}{M_W})^2
(1-\frac{4M_B^2}{M_A^{02}})^{1/2}, \nonumber\\
\Gamma(A^0\to T\bar T)= \frac{3g^2m_{A^0}}{32\pi}  \tan^2\beta
 (\frac{M_T}{M_W})^2
(1-\frac{4M_T^2}{M_A^{02}})^{1/2}.
\eeqn
These may be compared with the decays of $A^0$ into a 4-th sequential generation which are
\beqn
\Gamma(A^0\to e_4\bar e_4)= \frac{g^2m_{A^0}}{32\pi} \tan^2\beta (\frac{m_{e_4}}{M_W})^2
(1-\frac{4m_{e_4}^2}{M_A^{02}})^{1/2}, \nonumber\\
\Gamma(A^0\to d_4\bar d_4)= \frac{3g^2m_{A^0}}{32\pi}   \tan^2\beta  (\frac{m_{d_4}}{M_W})^2
(1-\frac{4m_{d_4}^2}{M_A^{02}})^{1/2}, \nonumber\\
\Gamma(A^0\to u_4\bar u_4)= \frac{3g^2m_{A^0}}{32\pi}  \cot^2\beta
 (\frac{m_{u_4}}{M_W})^2
(1-\frac{4m_{u_4}^2}{M_A^{02}})^{1/2}.
\eeqn
A study of the branching ratios will differentiate between a sequential fourth  generation and 
a mirror fourth generation.

\end{document}